\documentclass[final,5p,times,twocolumn,authoryear]{elsarticle}

\usepackage{amssymb}

\usepackage{amsmath,bm}
\usepackage{algorithmic}
\usepackage{listings}
\def\astrobj#1{#1}

\RequirePackage[l2tabu,orthobox]{nag}

\newcommand{\FSeq}{FSeq}
\newcommand{\SSeq}{SSeq}
\newcommand{\ISeq}{ISeq}
\newcommand{\fseq}{fast sequence}
\newcommand{\sseq}{slow sequence}
\newcommand{\iseq}{intermediate sequence}

\journal{New Astronomy}

\begin{document}

\begin{frontmatter}



\title{GOTHIC: Gravitational oct-tree code accelerated by hierarchical time step controlling}


\author[label1,label2]{Yohei Miki}
\author[label1,label2]{Masayuki Umemura}

\address[label1]{Center for Computational Sciences, University of Tsukuba, 1-1-1 Tennodai, Tsukuba, Ibaraki 305-8577, Japan}
\address[label2]{CREST, JST, 1-1-1 Tennodai, Tsukuba, Ibaraki 305-8577, Japan}

\begin{abstract}
The tree method is a widely implemented algorithm for collisionless $N$-body simulations in astrophysics well suited for GPU(s). 
Adopting hierarchical time stepping can accelerate $N$-body simulations; however, it is infrequently implemented and its potential remains untested in GPU implementations. 
We have developed a Gravitational Oct-Tree code accelerated by HIerarchical time step Controlling named \texttt{GOTHIC}, which adopts both the tree method and the hierarchical time step. 
The code adopts some adaptive optimizations by monitoring the execution time of each function on-the-fly and minimizes the time-to-solution by balancing the measured time of multiple functions. 
Results of performance measurements with realistic particle distribution performed on NVIDIA Tesla M2090, K20X, and GeForce GTX TITAN X, which are representative GPUs of the Fermi, Kepler, and Maxwell generation of GPUs, show that the hierarchical time step achieves a speedup by a factor of around 3--5 times compared to the shared time step. 
The measured elapsed time per step of \texttt{GOTHIC} is 0.30~s or 0.44~s on GTX TITAN X when the particle distribution represents the Andromeda galaxy or the NFW sphere, respectively, with $2^{24} =$~16,777,216 particles. 
The averaged performance of the code corresponds to 10--30\% of the theoretical single precision peak performance of the GPU. 
\end{abstract}


\end{frontmatter}


\section{Introduction}
\label{sec:intro}
Collisionless $N$-body simulations are frequently employed to investigate large scale structure formation and the formation and evolution of gravitational many-body systems such as galaxies. 
The acceleration of $N$-body particles is given by Newton's equation of motion,
\begin{equation}
  \bm{a}_i = \sum_{j=0, j \neq i}^{N-1} \frac{G m_j \left(\bm{r}_j - \bm{r}_i\right)}{\left(\left|\bm{r}_j - \bm{r}_i\right|^2 + \epsilon^2\right)^{3/2}},
\end{equation}
where $m_i$, $\bm{r}_i$, and $\bm{a}_i$ are the mass, position, and acceleration of the $i$-th particle of $N$ particles, respectively. 
The remaining symbols are the gravitational constant $G$ and the Plummer softening parameter $\epsilon$. 
The latter is commonly adopted in collisionless $N$-body simulations to eliminate divergence due to division by zero. 
Hereafter, we call the particles which feel and cause gravitational force as \textit{i-} and \textit{j-particles}, respectively, and denote their total numbers $N_i$ or $N_j$.

Employing a large number of $N$-body particles is essential for performing $N$-body simulations that resolve astrophysical phenomena.
Since the computational cost of order $O(N_i N_j)$ is too high to investigate realistic phenomena in detail, many earlier studies have attempted to accelerate $N$-body simulations. 
Widely used algorithms for reducing the amount of computations are the particle-mesh method and the tree method \citep{HockneyEastwood1988, BarnesHut1986}. 
The computational complexity of the tree method is $O(N_i \log{N_j})$ because the multipole expansion technique significantly reduces the contribution from \textit{j-}particles. 

Many $N$-body simulations adopt a shared time step, and that means all $N$-body particles share the time step that is required to track the orbital evolution of the particle that evolves its physical quantities in the shortest time span. 
The timescale of the evolution is not uniform in most astrophysical phenomena; for example, the free-fall time, which is a measure for the timescale of evolution due to self-gravity, scales as the inverse square root of the mass density, and the mass densities have order-of-magnitude differences in typical systems. 
Therefore, adopting a shared time step causes unnecessary, additional computations to track the evolution of the system. 
To overcome the situation, a scheme in which every $N$-body particle has their own individual time step was introduced by \citet{Aarseth1963}. 
Because individual time steps for all particles is not suitable for parallelization, \citet{McMillan1986} proposed the use of block time steps (or sometimes called hierarchical time steps) in which a group of particles has the same time step. 
Adopting block time steps can reduce the number of computations by reducing $N_i$. 

Exploiting accelerator devices is another approach to reducing the time-to-solution. 
In the field of numerical astrophysics, a famous accelerator for $N$-body simulations is the GRAPE (``GRAvity PipE'') series \citep{GRAPE, GRAPE1, GRAPE2, GRAPE2A, GRAPE1A, GRAPE6A, GRAPE3, GRAPE4, GRAPE6, GRAPE5, FIRST}. 
Its high performance is a result of the pipelined and massively parallel architecture design, which enables massive parallelization of gravitational force calculations. 
Another widely used accelerator device is the Graphics Processing Unit (GPU), which was originally developed as a processor dedicated to image processing, and is equipped with a large number of computing units (typically a few hundred to a few thousand), suitable for parallel computing. 
The memory architecture of GPU mainly consists of shared memory and global memory: the former is fast and small on-chip memory ($\sim$1~MB per GPU), and the latter is slow and large off-chip memory ($\sim$1--10~GB per GPU, but about 100 times slower than the shared memory). 
Rapid performance improvement of GPUs and the development of General Purpose computing on GPU (GPGPU) have elevated GPUs to be the most attractive accelerators. 
Moreover, recent demands for power efficient devices strongly support the rapid development of accelerator devices such as GRAPE, GPU, and Intel Xeon Phi. 

To promote GPU computing, NVIDIA provides the C/C++ like programming environment named Compute Unified Device Architecture \citep[CUDA:][]{CUDA1.0Manual, CUDA7.5Manual}. 
CUDA helps programmers implement GPU codes and optimize them by abstracting actual management of GPU cores and hiding differences among GPUs of various generations. 
For example, an essential building block of the Fermi generation GPUs is the streaming multiprocessor (SM), which is a group of 32 CUDA cores. 
In the Kepler generation of GPUs and Maxwell generation of GPUs SM are called SMX and SMM, respectively, and have 192 or 128 CUDA cores. 
For simplicity, we will refer to this fundamental group of CUDA cores as SM, irrespective of the GPU's generation. 
The fundamental parallelism in CUDA is thread parallelism, and a bunch of threads is called a block (typically 128--512 threads). 
Also, a group of blocks is called a grid; the hierarchical structure composed of the thread, block and grid is a key concept in CUDA. 
CUDA assigns multiple blocks to an SM for hiding latency to access memory and switch threads effectively. 
Since, in most applications, the number of threads per SM is sufficiently large compared to the number of CUDA cores per SM, all we have to do is to determine the number of threads per block. 
Through such abstractions of programming and the achieved high performance, GPU computing is now an important domain in high performance computing (HPC) community. 

Many earlier studies showed that the tree method efficiently works on GPU(s) \citep{Nakasato2012, Ogiya2013, Bedorf2012, Bedorf2014, WatanabeNakasato2014}. 
However, none of the studies have coupled their tree method with the block time step on GPU. 
One difficulty when coupling the block time step with the tree code running on GPU is maintaining performance in the low $N_i$-regime. 
As mentioned above, the reduction of the time-to-solution by the block time step is due to the decrease of $N_i$. 
However, the performance of massively parallel architectures always drops in the low-number limit because only some of the cores perform any computations while others do not, leading to a waste of computing resources. 
In the typical implementation of a direct $N$-body code running on GPU, the critical number of particles required in order not to waste CUDA cores is $10^4$ \citep{Miki2012, Miki2013}.
A viable method to decrease the critical number is to adopt \textit{ij}-parallelization \citep{Nitadori2006, Nyland2007, Miki2012}, by which multiple processors calculate the force on a common particle. 
\citet{Miki2012} showed that \textit{ij}-parallelization can sustain the high performance of their direct $N$-body code down to $N \sim 10^3$ on NVIDIA Tesla C2070. 
An option to activate \textit{ij}-parallelization may increase the performance of tree code on GPU that adopts the block time step. 

In GPU computing, a bunch of threads, 32 threads in the case of CUDA called a warp, always execute the same operation concurrently. 
If two threads in a warp are forced to execute different operations due to conditional branching, then the threads run both operations. 
Since there are 32 threads in a warp, this behavior, named ``warp divergence'', may cause up to 32 times slow down of calculations in the worst case. 
Therefore, avoiding the warp divergence is one of the key strategies to accelerate calculations using GPU. 
In the tree code runs on GPU, \citet{Ogiya2013} proposed an algorithm that reduces the warp divergence within the tree traversal and showed it improves the performance. 
On the other hand, concurrent operations by 32 threads present an opportunity to remove explicit synchronizations within a warp because they are implicitly synchronized. 
Synchronization is an inevitable operation for parallel computing to proceed properly; however, it often hinders achieving high performance. 
Hence, removing explicit synchronizations recovers high performance in parallel computing and reduces the time-to-solution. 
In $N$-body simulation with direct summation, \citet{Miki2012} demonstrated the benefits of removing explicit synchronizations, especially in the low $N$ runs, where the contribution from synchronization grows. 

There is further room for accelerating $N$-body simulations through automatic performance tuning (auto-tuning). 
Several examples of auto-tuning accelerating software libraries have been developed in the HPC community \citep[e.g.,][]{Whaley2001, FrigoJohnson2005}. 
The primary purpose of auto-tuning is to provide performance portability on various architectures and to benefit from the rapid performance improvements of architectures without needing to significantly modify optimized codes. 
Another essential objective of auto-tuning is to ensure the high performance of the code irrespective of input. 
For example, the performance of sparse matrix-vector multiplications (SpMV) on GPU has a strong dependence on the input sparse matrix \citep{BellGarland2008}. Many studies showed the benefits of auto-tuning for SpMV \citep{RegulyGiles2012, Ashari2014, LiuVinter2015, MaggioniBerger-Wolf2016}. 
In astrophysics, \citet{Ishiyama2009, Ishiyama2012} achieved a good load balance for their massively parallel TreePM code by incorporating on-the-fly measurements for the execution time of each function within the simulation. 
Just like SpMV, the time-to-solution of the tree method are dependent on the initial data because the particle distribution determines the total number of calculated interactions. 
Introducing some adaptive features to the tree code would contribute to accelerating $N$-body simulations by reducing slowdowns in the computation due to the non-uniform particle distribution. 

These considerations drove us to develop and test a tree code adopting a block time step that runs on GPU. 
The name of the code is \texttt{GOTHIC} (Gravitational Oct-Tree code accelerated by HIerarchical time step Controlling). 
The remainder of this paper is organized as follows. 
Section~\ref{sec:implementation} introduces the implementation and optimizations of \texttt{GOTHIC} using CUDA. 
Section~\ref{sec:results} presents results of performance measurements, and Section~\ref{sec:discussion} contains discussions. 
Finally, Section~\ref{sec:summary} summarizes this work. 

\section{Implementation}
\label{sec:implementation}
This section describes our strategy, implementation, and optimizations in detail. 
In \texttt{GOTHIC}, all instructions are performed on GPU, just like \texttt{Bonsai} \citep{Bedorf2012, Bedorf2014} to minimize communication between CPU and GPU. 
Also, all floating-point operations are performed in single precision because this provides sufficient accuracy to follow the time evolution of collisionless systems. 
Section~\ref{sec:implementation:makeTree} explains how to construct tree structure on GPU, and Section~\ref{sec:implementation:walkTree} presents the algorithm to calculate the gravitational force adopted in \texttt{GOTHIC}. 
Sections~\ref{sec:implementation:splitGroups} and \ref{sec:implementation:ij.parallelization} introduce additional optimizations aiming to keep performance even in situations not suitable for GPU. 
Section~\ref{sec:implementation:rebuildInterval} gives information on further optimization to reduce the time-to-solution of \texttt{GOTHIC}, rather the execution time of a specific function. 
Sections~\ref{sec:implementation:MAC} and \ref{sec:implementation:integrator} present other information required to implement a tree code, and Section~\ref{sec:implementation:kepler} shows additional tips and issues related to the Kepler generation GPUs. 

\subsection{Generating Tree Structure}
\label{sec:implementation:makeTree}
\begin{figure}
  \centering
  \includegraphics[width=.6\linewidth,clip]{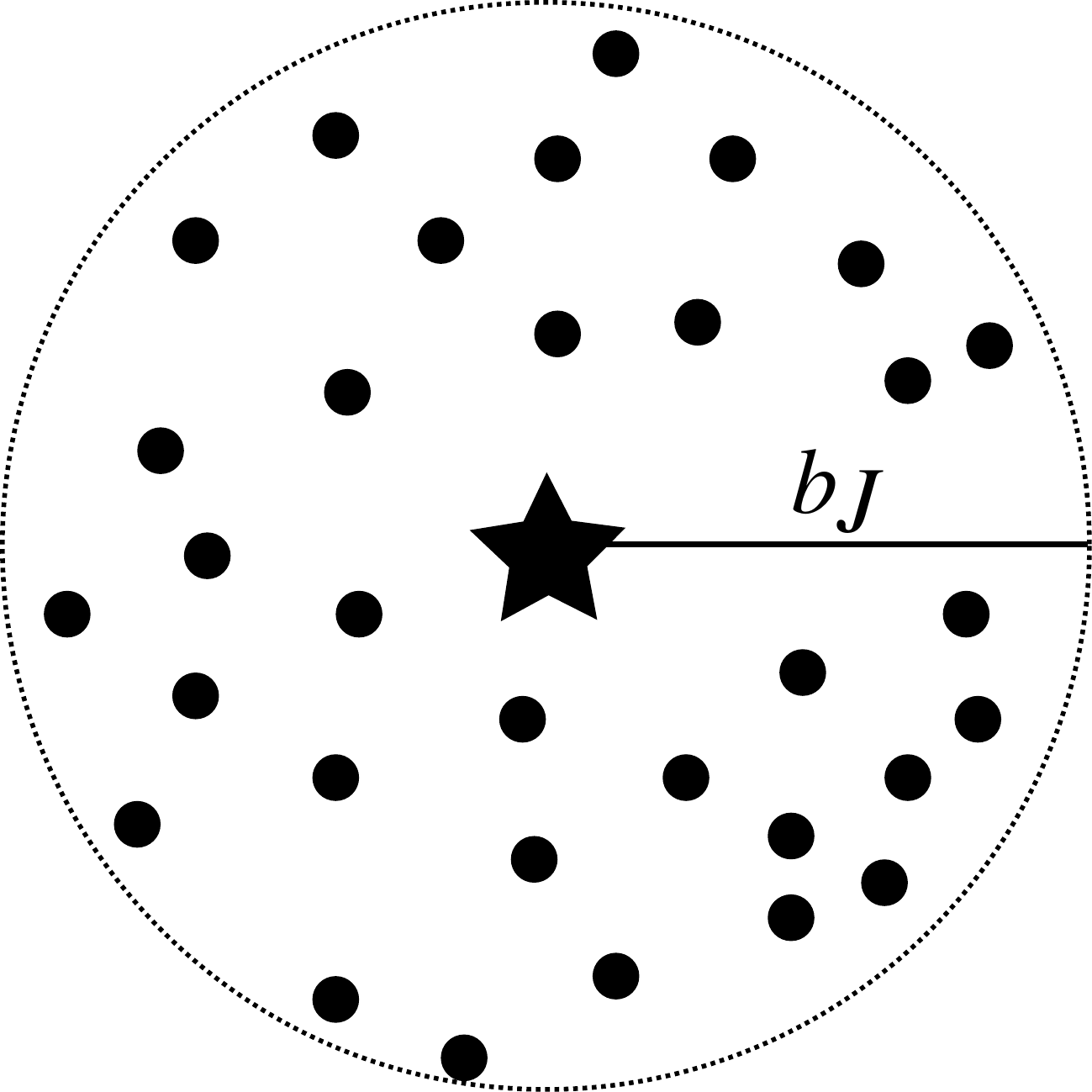}
  \caption{
    Definition of a pseudo \textit{j}-particle. 
    The filled circles and the star indicate locations of real $N$-body particles and the corresponding pseudo \textit{j}-particle, respectively. 
    The dotted circle represents the size of the pseudo \textit{j}-particle $b_J$. 
  }
  \label{fig:pseudo.j-particle}
\end{figure}
The space-filling curve based construction of the tree structure, which represents the particle distribution as a logical structure, is performed by the GPU. 
In this study, we adopt the Peano--Hilbert space-filling curve \citep{Sagan2012} to exploit its one-stroke sketch nature, which the more familiar Morton curve does not have. 
First, the GPU generates the Peano--Hilbert key for all $N$-body particles in the global memory of the device (see \ref{sec:appendix:space.filling.curves} for more details). 
Then, the $N$-body particles are sorted according to the Peano--Hilbert space-filling curve by using \texttt{cub::DeviceRadixSort::SortPairs} function provided in CUB\footnote{http://nvlabs.github.io/cub/index.html} v1.5.1. 
Using the Peano--Hilbert curve guarantees that the particles near one another in memory space are also near one another in physical space. 
The relation between memory space and physical space is important when optimizing codes, as shown by \citet{Ogiya2013} for accelerating gravity calculations using the tree structure. 

Next, the GPU links the Peano--Hilbert key with the tree structure. 
The Peano--Hilbert space-filling curve itself has a hierarchical structure. 
Dividing a cube into eight sub-cubes (i.e., generating an octree structure) corresponds to dividing the Peano--Hilbert key into eight equal parts (or finding seven partitions of the Peano--Hilbert key). 
Because increasing parallelism is essential to accelerating calculations using many-core architectures such as GPU, we construct the tree structure in a breadth-first manner. 
Checking multiple tree cells in parallel is possible. However, child cells of all checked cells must have serial numbers to identify them. 
Calculating prefix sums \citep{Blelloch1990} is necessary to tag all tree cells consistently. 

When calculating prefix sum within a warp in parallel, the implicit synchronization of 32 threads is an important feature to exploit. 
Since the warp shuffle instruction is available in GPUs starting with the Kepler generation, our implementation of parallel prefix sum calculation within a warp utilizes the warp shuffle instruction on the Kepler and Maxwell generation GPUs or the shared memory on the Fermi generation GPUs. 
Repeated executions of a parallel scan within a warp with the appropriate use of \texttt{\_\_syncthreads()} and shared memory yield parallel prefix sums within a block. 
To implement parallel prefix sums within a grid, global synchronization of multiple blocks within a grid is necessary. 
\texttt{GOTHIC} adopts the GPU lock-free synchronization proposed by \citet{XiaoFeng2010} as a global synchronization mechanism. 
In the algorithm, all blocks within a grid must run simultaneously so as not to cause a deadlock. 
The \texttt{\_\_launch\_bounds\_\_} qualifier is useful to control the number of concurrent blocks in the case that the register usage limits the number of concurrent blocks per SM.
Also, \texttt{cudaFuncAttributes::numRegs} obtained by calling the \texttt{cudaFuncGetAttributes} function is helpful to judge whether the deadlock will occur just before calling the device function. 

Since \texttt{GOTHIC} adopts the monopole approximation for gravity calculation between an \textit{i}-particle with a tree cell, introducing imaginary particles corresponding to tree cells can simplify the implementation of the function to calculate the gravitational force. 
After the Peano--Hilbert keys are associated with the tree structure, the GPU generates imaginary particles called pseudo \textit{j}-particles and connect them with tree cells. 
The pseudo \textit{j}-particle has information on mass $m_J$, position $\bm{r}_J$ and the size $b_J$; hereafter, the capitalized subscript indicates the index of the pseudo particles. 
The mass is the total mass of real $N$-body particles contained in the corresponding tree cell and the position is the center-of-mass of the particles. 
The size of the pseudo \textit{j}-particle is defined as the radius of a sphere centered on $\bm{r}_J$ which can contain all $N$-body particles contained in the tree cell (see Fig.~\ref{fig:pseudo.j-particle}). 
All physical quantities of the pseudo \textit{j}-particles must be recalculated at every time step to calculate gravitational force properly. 

\subsection{Multipole Acceptance Criterion}
\label{sec:implementation:MAC}
If a pseudo \textit{j}-particle is far, then the gravity from the particle is calculated; if it is near, the tree cell is restricted to the lower level.
To judge whether a pseudo \textit{j}-particle is near or far, the Multipole Acceptance Criterion (MAC) is employed. The most simple MAC is opening angle criterion proposed by \citet{BarnesHut1986}:
\begin{equation}
  \frac{b_J}{d_{iJ}} \leq \theta,
  \label{eq:implementation:MAC:BH86}
\end{equation}
where $d_{iJ}$ is the distance to the particle from an \textit{i}-particle and $\theta$ is an accuracy controlling parameter. 

Because the above MAC cannot directly control the accuracy with which the gravitational forces are calculated, more sophisticated MACs have been proposed. 
The MAC proposed by \citet{WarrenSalmon1993, SalmonWarren1994} is as follows: 
\begin{equation}
  d_{iJ} \geq \frac{b_J}{2} + \sqrt{\frac{b_J^2}{4} + \sqrt{\frac{3 B_2}{\Delta_\mathrm{mul}}}},
  \label{eq:implementation:MAC:WS93}
\end{equation}
where $\Delta_\mathrm{mul}$ is an accuracy controlling parameter and 
\begin{equation}
  B_2 \equiv \sum_j m_j \left(\bm{r}_j - \bm{r}_J\right)^2. 
\end{equation}
The MAC defined by Eq.~(\ref{eq:implementation:MAC:WS93}) ensures the required accuracy by monitoring the truncation error of the multipole expansion. 

In addition, the acceleration MAC by \citet{Springel2005} given by
\begin{equation}
  d_{iJ} \geq \left(\frac{G m_J {b_J}^2}{\Delta_\mathrm{acc} \left|\bm{a}_i^\mathrm{old}\right|}\right)^{1/4}
  \label{eq:implementation:MAC:S05}
\end{equation}
also gives the required accuracy, where $\bm{a}_i^\mathrm{old}$ is the acceleration of the \textit{i}-particle in the previous time step and $\Delta_\mathrm{acc}$ is an accuracy controlling parameter. 
This MAC directly monitors the acceleration of each \textit{i}-particle, and gives the appropriate accuracy of the acceleration specified by $\Delta_\mathrm{acc}$. 

The best choice of MAC from the three above must be determined by experiments. 
In the case of a tree code running on CPU, \citet{Nelson2009} compared the elapsed time of each MAC as a function of achieved accuracy, and concluded that the acceleration MAC was the optimal choice. 
The performance of the MAC, however, should depend on the implementation of the function which calculates the gravitational acceleration and is optimized for a specific architecture, in our case, GPU. 
Comparing MACs is, therefore, still necessary for tree codes optimized for GPU and we will provide results of the comparison in \S\ref{sec:results:time}. 

\subsection{Traversing Tree Structure}
\label{sec:implementation:walkTree}
\begin{figure}
  \centering
  \includegraphics[width=.99\linewidth,clip]{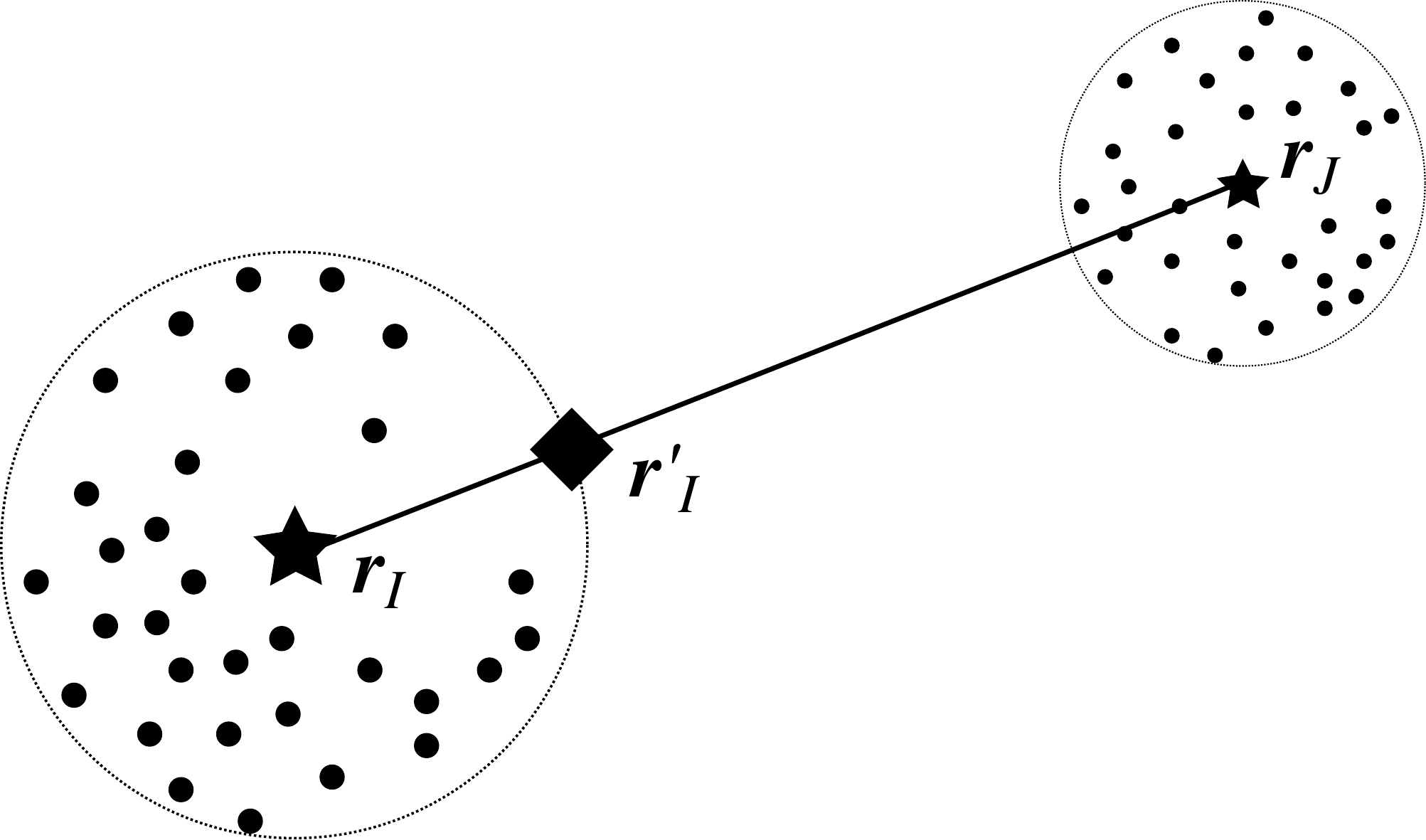}
  \caption{
    Sketch of the distance evaluation between \textit{i}-particles and a pseudo \textit{j}-particle. 
    Filled stars show positions of a pseudo \textit{i}-particle ($\bm{r}_I$) and a pseudo \textit{j}-particle ($\bm{r}_J$). 
    Filled circles enclosed by a dotted circle centered on the pseudo \textit{i}-particle are real \textit{i}-particles. 
    The filled diamond shows the possible nearest position of \textit{i}-particles to the pseudo \textit{j}-particle, $\bm{r}'_I$. 
    The distance between the pseudo \textit{i}-particle and pseudo \textit{j}-particle is measured as $|\bm{r}_J - \bm{r}'_I|$. 
  }
  \label{fig:pseudo.i-particle}
\end{figure}
Increasing arithmetic intensity leads to performance improvements since hiding the latency to access global memory becomes much easier. 
To increase the arithmetic intensity in the kernel function, \citet{Ogiya2013} introduced the technique of ``vectorization''. 
\citet{Ogiya2013} adopted the depth-first search on-the-fly and the number of \textit{i}-particles per thread is assumed to be $N_\mathrm{vec} (\geq 1)$. 
When judging whether the distance to a pseudo \textit{j}-particle is far or near, they calculate the distance between the pseudo \textit{j}-particle and $N_\mathrm{vec}$ \textit{i}-particles one by one. 
A minimum of $N_\mathrm{vec}$ evaluations of distance is used for the distance judgment. 
The total number of interactions increases due to the minimum of $N_\mathrm{vec}$ evaluations; therefore, $N_\mathrm{vec}$ has some optimal value determined by balancing pros and cons of the effects by the vectorization. 

During tree traversal when calculating the gravitational force, warp divergence occurs when some threads in a warp judge the distance to a pseudo \textit{j}-particle to be sufficiently far while the remainder judge the distance to still be near. 
\citet{Ogiya2013} introduced ``grouping'' to reduce the warp divergence. 
In this step, they group the distance judgment into $N_\mathrm{grp}$ threads ($N_\mathrm{grp}$ must be smaller than 32 to utilize the implicit synchronization within a warp) by sharing the minimum distance to a pseudo \textit{j}-particle from $N_\mathrm{vec}$ \textit{i}-particles in $N_\mathrm{grp}$ threads. 
Just like $N_\mathrm{vec}$, there is also an optimal value of $N_\mathrm{grp}$. 

In \citet{Ogiya2013}, $N_\mathrm{vec}$ distance calculations by $N_\mathrm{grp}$ threads and $\log_2{N_\mathrm{grp}}$ comparisons to group the judgement in $N_\mathrm{grp}$ threads are required to judge whether a specific pseudo \textit{j}-particle is near or far. 
Here, we modify the vectorization method proposed by \citet{Ogiya2013} by using a breadth-first search. 
The vectorization in the original form requires $N_\mathrm{vec}$-times $N_i$ particles for sustained performance. 
This diminishes the benefits of the block time step. 
Therefore, we adopt a compromise between an on-the-fly method and an interaction list method. 
In this method, a small sized interaction list is created in shared memory. 
Once the size of the interaction list reaches a certain predefined value, we calculate gravitational forces between \textit{i}-particles and pseudo \textit{j}-particles in the list and clear the list. 
By repeating the procedure, the gravity by all \textit{j}-particles is properly calculated. 
The arithmetic intensity of the kernel function is determined by the capacity of the interaction list, which depends on the number of threads per block and the cache configuration of the shared memory. 

To use the shared memory efficiently and reduce warp divergence, we adopt the grouping almost in the original form. 
Grouping the interaction list of $N_\mathrm{grp}$ threads leads to a $N_\mathrm{grp}$ times bigger list to be stored in the shared memory. 
We modify the algorithm for grouping the distance judgment to remove $\log_2{N_\mathrm{grp}}$ comparisons as follows. 
Since the breadth-first search can access queued tree cells in parallel, distance evaluations to multiple pseudo \textit{j}-particles can be performed at the same time. 
We introduce a pseudo \textit{i}-particle shared by $N_\mathrm{grp}$ threads as shown in Fig.~\ref{fig:pseudo.i-particle}. 
The pseudo \textit{i}-particles is to include all corresponding real \textit{i}-particles by defining the appropriate radius $b_I$. 
There is some freedom in defining the center of the enclosing sphere $\bm{r}_I$: for example, the center of the smallest enclosing ball, the center-of-mass of real \textit{i}-particles, or the geometric center of the enclosing rectangular cuboid (see \ref{sec:appendix:enclosing.balls} for more detail). 
The optimal choice to minimize the elapsed time of \texttt{GOTHIC} will be determined in \S\ref{sec:results:configuration} to provide the shortest elapsed time in micro-benchmarks. 
The distance between the pseudo \textit{i}-particle and a pseudo \textit{j}-particle is evaluated as the distance between an imaginary particle and the pseudo \textit{j}-particle. 
The imaginary particle is set at the intersection of the surface of the pseudo \textit{i}-particle with a line connects the pseudo \textit{i}-particle with the pseudo \textit{j}-particle, $\bm{r}'_I$: 
\begin{equation}
  |\bm{r}_J - \bm{r}'_I| \equiv \lambda r_{JI},
\end{equation}
where 
\begin{equation}
  \lambda =
  \begin{cases}
  1 - \frac{b_I}{r_{JI}} & (b_I < r_{JI}),\\
  0 & (b_I \geq r_{JI}).
  \end{cases}
\end{equation}
Introducing the pseudo \textit{i}-particle is functionally the same as the vectorization and the grouping by \citet{Ogiya2013} because the distance between the pseudo \textit{i}-particle and the \textit{j}-particle is always smaller than that between all corresponding \textit{i}-particles and the \textit{j}-particle. 

When traversing the tree structure in a breadth-first manner, many tree cells must be stored in a large buffer compared to one child cell stored under the depth-first search. 
The breadth-first search requires additional global memory allocation.
Because the total capacity of the global memory on GPU is limited (e.g., 5 GB for NVIDIA Tesla M2090 and K20X with ECC enabled), sophisticated memory management is necessary. 
In order to allocate as large as possible a chunk of global memory for the buffer, we first query the unused capacity of the global memory using the \texttt{cudaMemGetInfo()} function and then allocate the buffer in the global memory. 
The next problem is the assignment of the buffer to each thread-block. 
In this study, the capacity of the shared memory sets the upper limit on the number of thread-blocks per SM to two. 
It determines the maximum number of thread-blocks which can run simultaneously, and we equally divide the buffer into the given number of pieces. 
The special register \texttt{\%smid} acquired by the inline PTX function tells the ID of SMs, and is useful to assign unused parts of the buffer to a running thread-block. 
It should be noted that \texttt{\%smid} is a volatile variable. 
Thus, a careful treatment is required to occupy and release the partitioned buffer correctly. 

\subsection{Splitting Particle Groups in Low Dense Region}
\label{sec:implementation:splitGroups}
One of the shortcomings of the method introduced in \S\ref{sec:implementation:walkTree} is an over-computation when \textit{i}-particles in a low dense region are selected as a group of \textit{i}-particles. 
To avoid this situation, we introduce a critical separation $r_\mathrm{crit}$ to judge whether to unify \textit{i}-particles into a group or not. 
If the value is too large or too small, then the elapsed time will become longer due to over-computation or over-splitting of the kernel, respectively. 
The critical separation $r_\mathrm{crit}$ must be set carefully to minimize the elapsed time; however, it is impossible to determine the optimal value before the calculation because $r_\mathrm{crit}$ depends on the particle distribution which evolves in the simulation. 
This leads us to set $r_\mathrm{crit}$ through trial-and-error during the simulation. 
In other words, we apply auto-tuning to determine the optimal value of $r_\mathrm{crit}$. 
The strategy we adopt is to search for the optimal $r_\mathrm{crit}$ by minimizing the GPU time to calculate gravity using Brent's method \citep{Press2007} and treating the GPU time as a function of $r_\mathrm{crit}$. 
Since the optimal value of $r_\mathrm{crit}$ would also depend on time, some perturbation on $r_\mathrm{crit}$ is additionally introduced. 
According to this scheme, $r_\mathrm{crit}$ automatically evolves to reduce the elapsed time. 

\subsection{Increasing Parallelism in Gravity Calculation}
\label{sec:implementation:ij.parallelization}
Maintaining the high performance of the code down to the low $N_i$-regime is an essential point to achieve high performance with the block time step. 
However, this is difficult because a lack of parallelism reduces the GPU performance by wasting CUDA cores. 
The critical number of particles to saturate GPU performance is $10^4$ in the case of direct $N$-body calculation \citep{Miki2012, Miki2013}. 
Some remedy should be introduced to limit the performance decrease in low $N_i$-regime. 
A straightforward remedy is introducing \textit{ij}-parallelization to increase parallelism \citep{Nitadori2006, Nyland2007, Miki2012}. 
In the case of \textit{ij}-parallelization, multiple threads share an \textit{i}-particle and calculate gravity to the particle. 
As a result, we regain running CUDA cores and GPU performance even in the low $N_i$-regime.

Introducing \textit{ij}-parallelization requires an implementation of a force accumulation process among multiple threads that share a common \textit{i}-particle. In this work, we have implemented an essentially identical version of the algorithm proposed by \citet{Miki2012}. 
In principle, either synchronization or exclusive control or both are inevitable to sum up the threads' results, and this always impedes the performance improvement in parallel computing. 
\citet{Miki2012} proposed an algorithm specialized for GPU computing to alleviate the burden of the force accumulation process. 
They remove explicit synchronization of multiple threads by aggressively utilizing the specification of CUDA that 32 threads in a warp always perform the same operation (implicit synchronization). 
Therefore, the number of threads that share an \textit{i}-particle, $S$, must satisfy $S \leq 32$. 

\subsection{Tree Rebuild Interval}
\label{sec:implementation:rebuildInterval}
The cost of tree construction, $t_\mathrm{make}$, is not negligibly small compared to that of tree traversal, $t_\mathrm{walk}$, and there is no requirement to rebuild the tree structure every time step. 
Since the particle distribution is almost the same for two time steps in succession, reusing the old tree structure will not deteriorate $t_\mathrm{walk}$ without additional cost to rebuild the tree structure. 
The mismatch between the tree structure and the actual particle distribution would increase the execution time, and the timescale of the increase is a function of the time evolution of the particle distribution. 
There ought to be an optimal interval to rebuild the tree structure and finding it is a task suited to auto-tuning. 

The code determines the rebuild interval $n$ by guessing the total elapsed time $t_\mathrm{tot}$. 
The total elapsed time between the tree constructions is given by
\begin{equation}
  t_\mathrm{tot} = t_\mathrm{make} + \sum_{i=1}^{n} t_\mathrm{walk}^{(i)},
  \label{eq:implementation:rebuildInterval:total.elapsed.time}
\end{equation}
where $t_\mathrm{walk}^{(i)}$ is the execution time to calculate gravity in the $i$-th step out of $n$ steps which use the same tree structure. 

Here, we introduce three toy models, a linear growth model, a power-law growth model, and a parabolic growth model, to guess $t_\mathrm{walk}^{(i)}$. 
In the first model, we assume $t_\mathrm{walk}$ grows as 
\begin{equation}
  t_\mathrm{walk}^{(i)} = t_1 + (i - 1) \varDelta t,
\end{equation}
where $t_1$ and $\varDelta t$ are intercept and slope, respectively. 
The above fitting parameters are determined using the least squared method by monitoring the execution time in every time steps. 
Then, the total elapsed time is estimated as 
\begin{equation}
  t_\mathrm{tot} = t_\mathrm{make} + n (t_1 - \varDelta t) + \frac{n (n + 1)}{2} \varDelta t.
  \label{eq:implementation:rebuildInterval:total.elapsed.time.in.linear.growth.model}
\end{equation}
To minimize the mean execution time $t_\mathrm{mean} \equiv t_\mathrm{tot} / n$, differentiate $t_\mathrm{mean}$ with respect to $n$: 
\begin{align}
  \frac{d}{dn} \frac{t_\mathrm{tot}}{n} &
  = - \frac{t_\mathrm{make}}{n^2} + \frac{\varDelta t}{2}.
\end{align}
Therefore, the condition to get the extremum is 
\begin{equation}
  n^2 = \frac{2}{\varDelta t} t_\mathrm{make}.
  \label{eq:implementation:rebuildInterval:optimal.condition.in.linear.growth.model}
\end{equation}
Furthermore, the second derivative with respect to $n$ is evaluated as 
\begin{equation}
  \frac{d^2}{dn^2} \frac{t_\mathrm{tot}}{n} = \frac{2 t_\mathrm{make}}{n^3},
\end{equation}
and is always positive meaning that Eq.~(\ref{eq:implementation:rebuildInterval:optimal.condition.in.linear.growth.model}) always minimizes the mean execution time. 

The power-law and the parabolic growth models are shown in \ref{sec:appendix:rebuild.interval.modeling}. 
The model which gives the smallest reduced $\chi^2$ value, 
\begin{equation}
  \chi^2_\nu \equiv \frac{1}{\nu} \sum_{i = 1}^n \left(\frac{t_\mathrm{walk, model}^{(i)} - t_\mathrm{walk, measured}^{(i)}}{\sigma_i}\right)^2,
\end{equation}
is the most appropriate of the three choices. 
The degrees of freedom $\nu$ is $n - 2$ (for the linear or power-law growth model) or $n - 3$ (for the parabolic growth model), and we simply assume $\sigma_i$ is unity. 

\subsection{Orbit Integration}
\label{sec:implementation:integrator}
When the block time step is employed, every \textit{i}-particle has its own time step.
Since the adaptive, block time step is employed, we adopt a second-order Runge-Kutta method to integrate the particle orbit. 
In the prediction step, we update positions and velocities of all \textit{j}-particles by 
\begin{equation}
  \bm{v}_j^{n + 1/2} = \bm{v}_j^n + \frac{\Delta t_j^n}{2} \bm{a}_j^n,
\end{equation}
\begin{equation}
  \bm{r}_j^{n + 1} = \bm{r}_j^n + \Delta t_j^n \bm{v}_j^{n + 1/2},
\end{equation}
where $\bm{v}_j^n$ is the velocity of the \textit{j}-th particle at the $n$-th time step, subscripts and superscripts indicate the index of particles and time step, respectively. 
We then calculate gravity from all \textit{j}-particles to selected \textit{i}-particles, and execute the correction step for the chosen \textit{i}-particles as 
\begin{equation}
  \bm{v}_i^{n + 1} = \bm{v}_i^{n + 1/2} + \frac{\Delta t_i^n}{2} \bm{a}_i^{n + 1}.
\end{equation}
Because the above predictor--corrector method is not a symplectic integrator, it does not conserve the pseudo-Hamiltonian unlike the leap-frog method often employed with the shared, fixed time step. 

For the comparison cases where the time step is shared and fixed, we adopt a second-order leap-frog method. 
In this case, orbit integration is performed as 
\begin{equation}
  \bm{v}_j^{n + 1/2} = \bm{v}_j^{n - 1/2} + \frac{\Delta t_j^n}{2} \bm{a}_j^n,
\end{equation}
\begin{equation}
  \bm{r}_j^{n + 1} = \bm{r}_j^n + \Delta t_j^n \bm{v}_j^{n + 1/2}.
\end{equation}
For fixed shared timesteps, the Runge-Kutta integrator reduces to the leap-frog method.

\subsection{Note for Kepler generation GPUs}
\label{sec:implementation:kepler}
Kepler generation GPUs support more functions that are useful in performance optimization compared to Fermi generation GPUs. 
One is warp shuffle instructions, which enable reading registers in other threads within a warp without using the shared memory. 
Warp shuffle instructions are heavily exploited in the calculation of parallel prefix sums and reductions since it is faster than accessing registers via shared memory. 
The read-only data cache is another feature to be noted. 
Just adding the \texttt{const \_\_restrict\_\_} qualifier tells the compiler to use a distinct cache in addition to L2 cache of the global memory. 
It effectively enlarges the capacity of cache and increases effective memory bandwidth. 

A warp scheduler has two instruction dispatch units (IDUs) on Kepler generation GPUs \citep{GK110Whitepaper} while it has only one IDU on Fermi generation GPUs \citep{GF110Whitepaper}. 
The presence of multiple IDUs within a warp scheduler causes scheduling issues if subsequent operations within a warp have mutual dependencies. 
Furthermore, \citet{LaiSeznec2013} reported that bank conflict of registers could occur among four banks on Kepler generation GPUs. 
Both the scheduling issue and bank conflict of registers cause a slowdown of operations on Kepler generation GPUs. 
Introducing instruction level parallelism can remove the dependency between subsequent operations and remove the scheduling issue created by multiple IDUs. 
We examined effects of increasing instruction level parallelism of multiple executions of fused multiply-add (FMA) instructions and direct $N$-body code without removing bank conflict of registers on NVIDIA Tesla K20. 
However, the performance did not improve; this suggests that a careful arrangement of registers to prevent bank conflicts is also necessary for further optimization. 
Because NVIDIA does not provide any tool or framework to arrange registers manually and CUDA shuffles locations of registers, we did not increase instruction level parallelism of subsequent computations or arrange locations of registers. 
Once such problems originated by hardware are resolved, we can expect the performance of \texttt{GOTHIC} to increase on Kepler or Maxwell generation GPUs. 

\section{Performance Measurements of the Code}
\label{sec:results}

\subsection{Configuration of Measurements}
\label{sec:results:configuration}
\begin{table*}
  \caption{Measurement Environment}
  \label{tab:results:configuration:environment}
  \centering
  \begin{tabular}{c|ccc}
    \hline\hline
    System & HA-PACS/BC & HA-PACS/TCA & Workstation\\
    \hline
    Number of nodes & 268 & 64 & 1\\
    CPU & Intel Xeon E5-2670 & Intel Xeon E5-2680 v2 & Intel Xeon E5-2640 v3\\
        & 8 cores, 2.6 GHz & 10 cores, 2.8 GHz & 8 cores, 2.6 GHz\\
        & \multicolumn{2}{c}{2 sockets per node} & 2 sockets\\
    RAM & DDR3-1600, 8 channels & DDR3-1866, 8 channels & DDR4-2133, 8 channels\\
        & \multicolumn{2}{c}{128 GB per node} & 64 GB\\
    GPU & NVIDIA Tesla M2090 & NVIDIA Tesla K20X & NVIDIA GeForce GTX TITAN X\\
        & 512 cores, 1.3 GHz & 2688 cores, 732 MHz & 3072 cores, 1 GHz\\
        & \multicolumn{2}{c}{4 boards per node} & 2 boards\\
    Video RAM & \multicolumn{2}{c}{6 GB (GDDR5, ECC on) per GPU} & 12 GB (GDDR5) per GPU\\
    C Compiler & \multicolumn{2}{c}{icc 15.0.5.223 (gcc 4.4.7 compatibility)} & gcc 4.8.5\\
    CUDA Toolkit & \multicolumn{3}{c}{7.5.17}\\
    \hline
  \end{tabular}
\end{table*}
Performance measurements were done on HA-PACS (Highly Accelerated Parallel Advanced system for Computational Sciences) and a workstation at the University of Tsukuba. 
HA-PACS is composed of two clusters: the Base Cluster (BC) and the Tightly Coupled Accelerator (TCA). 
HA-PACS/BC and HA-PACS/TCA is equipped with NVIDIA Tesla M2090 (Fermi generation GPU) and NVIDIA Tesla K20X (Kepler generation GPU), respectively. 
NVIDIA GeForce GTX TITAN X (Maxwell generation GPU) is installed on the workstation. 
Table~\ref{tab:results:configuration:environment} lists the detailed information of the measurement environments. 
All environments have multiple GPUs, but we use only a single board of GPU on each machine in the measurements below. 

\begin{table}
  \caption{Configuration of thread-blocks}
  \label{tab:results:configuration:kernel}
  \centering
  \begin{tabular}{c|c||rrr|r}
    \hline\hline
    function & GPU$^{(a)}$ & $T_\mathrm{tot}$$^{(b)}$ & $T_\mathrm{sub}$$^{(c)}$ & $S$$^{(d)}$ & $R$$^{(e)}$\\
    \hline
    \texttt{walkTree} & M2090   &  256 & 32 &  1 & 63\\
                      & K20X    &  512 & 32 &  1 & 64\\
                      & TITAN X &  512 & 32 &  4 & 64\\
    \texttt{makeTree} & M2090   &  128 &  8 &    & 53\\
                      & K20X    &  128 &  8 &    & 49\\
                      & TITAN X &  128 &  8 &    & 64\\
    \texttt{linkTree} & M2090   &  128 &    &    & 27\\
                      & K20X    &  256 &    &    & 27\\
                      & TITAN X &  256 &    &    & 23\\
    \texttt{trimTree} & M2090   &  128 &    &    & 18\\
                      & K20X    &  128 &    &    & 22\\
                      & TITAN X &  128 &    &    & 22\\
    \texttt{calcMAC}  & M2090   &  128 & 32 &    & 59\\
                      & K20X    &  128 & 32 &    & 55\\
                      & TITAN X &  256 & 32 &    & 64\\
    \texttt{genPHkey} & M2090   &  256 &    &    & 36\\
                      & K20X    & 1024 &    &    & 40\\
                      & TITAN X & 1024 &    &    & 40\\
    \hline
  \end{tabular}
  \begin{itemize}
    \setlength{\itemsep}{-2pt}
    \item[(a)] Name of GPU. 
    \item[(b)] Number of threads per block. 
    \item[(c)] Number of threads share operations. 
    \item[(d)] Number of threads share an \textit{i}-particle. 
    \item[(e)] Register usage per thread. 
  \end{itemize}
\end{table}
Fundamental parameters of the code (e.g., the number of threads per block for each kernel function) are determined by micro-benchmarks for a Navarro--Frenk--White (NFW) sphere \citep{Navarro1995, Navarro1996}, a Plummer sphere \citep{Plummer1911}, a King sphere \citep{Michie1963, MichieBodenheimer1963, King1966} and a Hernquist sphere \citep{Hernquist1990}. 
All initial conditions used in this study are generated by the MAny-component Galactic Initial-conditions (MAGI) generator \citep{MAGI}. 
Table~\ref{tab:results:configuration:kernel} summarizes the resultant configuration for functions related to the tree structure. 
Obviously, optimal values exist for each function (\texttt{walkTree} executes tree traversal, \texttt{makeTree}, \texttt{linkTree}, and \texttt{trimTree} build tree structure, \texttt{calcMAC} calculates physical quantities of pseudo \textit{j}-particles, and \texttt{genPHkey} translates the position of an \textit{i}-particle to a Peano--Hilbert key). 
The adopted enclosing ball for \texttt{walkTree} is the efficient bounding sphere \citep{Ritter1990} on GTX TITAN X, while M2090 and K20X use the sphere centered on the geometric center of the enclosing rectangular cuboid. 

\subsection{Measured Elapsed Time}
\label{sec:results:time}
\begin{figure*}
  \centering
  \includegraphics[viewport=138 153 821 616, width=.99\linewidth,clip]{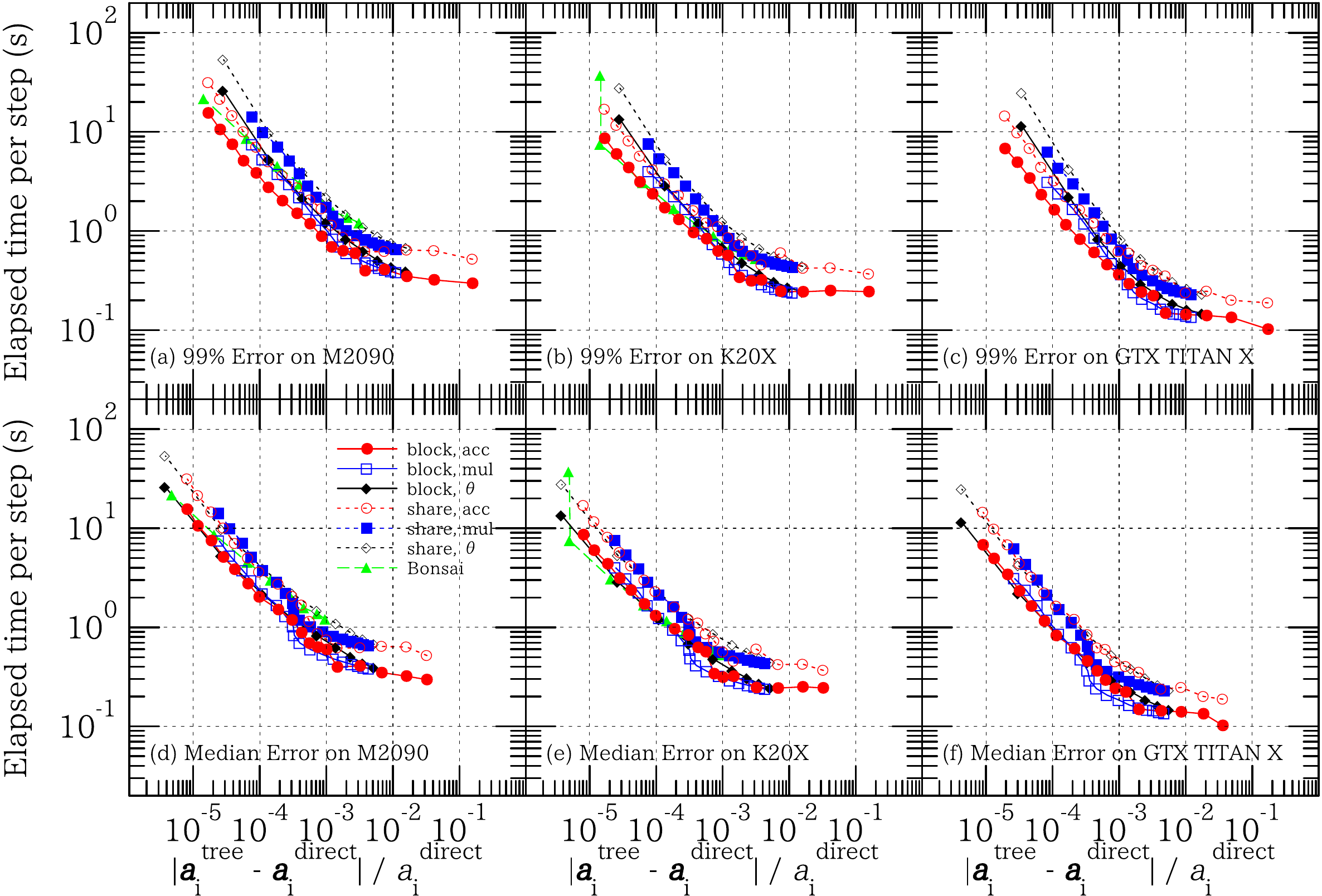}
  \caption{
    Elapsed time per step as a function of force accuracy. 
    Distribution of the $N$-body particles is an NFW sphere with $2^{23} =$~8,388,608 particles. 
    Solid and dotted lines with symbols are results of the block time step and shared time step, respectively. 
    Each symbol indicates different MACs: red circles are acceleration MAC \citep{Springel2005}, blue squares are multipole MAC \citep{WarrenSalmon1993}, and black diamonds are opening angle \citep{BarnesHut1986}. 
    The green triangles with dashed line show the elapsed time of the public code \texttt{Bonsai} \citep{Bedorf2012, Bedorf2014}. 
    Values of the accuracy controlling parameters are $2^{-2}$, $2^{-3}$, $\cdots$, $2^{-19}$ for the acceleration MAC and the multipole MAC, $0.9$, $0.8$, $\cdots$, $0.1$ for the opening angle and \texttt{Bonsai} from right to left. 
    Upper and lower panels show the measured elapsed time against 99\% error and median error of acceleration as a vector, respectively. 
    Each panel exhibits benchmark results on different GPUs: left (M2090), middle (K20X), and right (GTX TITAN X). 
  }
  \label{fig:elapsed.time@NFW}
\end{figure*}
First, we investigated relations among the accuracy controlling parameters of three MACs (\S\ref{sec:implementation:MAC}), the resultant accuracy of gravity calculation and the elapsed time on various generations of GPUs. 
This is similar to the evaluation of a tree code performed by \citet{Nelson2009}. 
Figure~\ref{fig:elapsed.time@NFW} shows the result in the case of an NFW sphere with $2^{23} =$~8,388,608 particles. 
The cutoff radii of the density profile and the length of the Plummer softening are $5 r_s$ and $r_s / 64$, respectively, where the scale length $r_s$ is set to unity. 
The elapsed time is evaluated as the wall clock time per time step (total number of time steps is fixed to 1024) to include the effects of auto-tuning; it also includes the time required to read/write files and allocate/deallocate memory. 
The accuracy of the gravity calculation is evaluated as a relative error of acceleration in the tree code $\bm{a}_i^\mathrm{tree}$ compared to acceleration in the direct $N$-body code, $\bm{a}_i^\mathrm{direct}$, where the subscript $i$ indicates the index of the $N$-body particles. 
Upper and lower panels of the figure present the results for the 99 percentile error and median error, respectively. 
In other words, the points trace the loci at which 99\% (50\%) of $N$-body particles have a smaller error of the acceleration than the plotted value for each MAC in the upper (lower) panels. 
The figure clearly reveals the block time step (solid lines) is roughly twice as fast as the shared time step (dotted lines). 
The block time step with the acceleration MAC (red filled circles with solid line) has the shortest elapsed time in most cases. 
The multipole MAC (blue squares) is sometimes the optimal choice, especially with lower accuracy, and its performance with higher accuracy is comparable to that of the opening angle (black diamonds). 

We have also compared the performance of \texttt{GOTHIC} with the public code \texttt{Bonsai}\footnote{https://github.com/treecode/Bonsai} \citep[][green triangles]{Bedorf2012, Bedorf2014} which runs on the Fermi and Kepler generation GPUs. 
On M2090, the performance measurement with $\theta = 0.1$ for \texttt{Bonsai} was not completed because the computation time was too long. 
In all cases, \texttt{GOTHIC} with acceleration MAC and block time step (i.e., fastest configuration) was faster than \texttt{Bonsai} except for the case for which the median force error was less than $\sim 10^{-5}$ on K20X. 
The figure clearly shows that the improvements of \texttt{GOTHIC} with respect to \texttt{Bonsai} are more significant on M2090 compared to K20X. 
This was expected, because \citet{Bedorf2014} performed sophisticated optimizations focused on the Kepler generation GPUs while we omit some optimizations (see \S\ref{sec:implementation:kepler}). 
In other words, the performance improvements of \texttt{GOTHIC} from \texttt{Bonsai} on the Kepler generation of GPUs would increase if we introduced optimizations highly focused on the Kepler generation of GPUs. 
The typical accuracy for $N$-body simulations on galactic scales is around $10^{-3}$, which  corresponds to a value of $\theta$ of 0.5--0.7. 
In such realistic parameter regions, \texttt{GOTHIC} is a few time faster than \texttt{Bonsai} on M2090. 

\begin{figure*}
  \centering
  \includegraphics[viewport=138 153 821 616, width=.99\linewidth,clip]{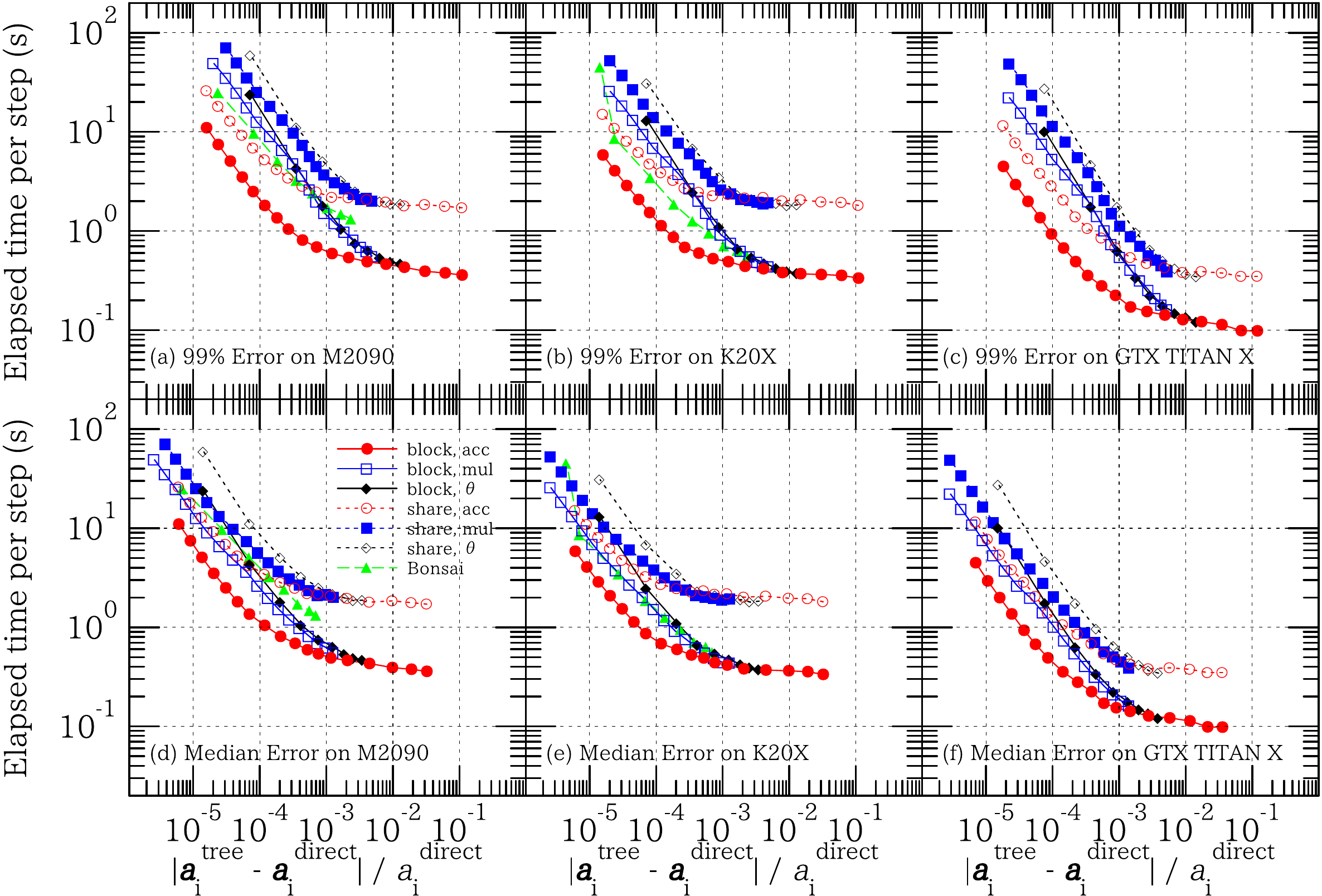}
  \caption{
    Elapsed time per step as a function of force accuracy. 
    The distribution of the $N$-body particles represents the spiral galaxy \astrobj{M31} with $2^{23} =$~8,388,608 particles. 
    Symbols, lines and panels are the same as those in Fig.~\ref{fig:elapsed.time@NFW}. 
  }
  \label{fig:elapsed.time@M31}
\end{figure*}
The NFW sphere is not suitable for evaluating effects of the block time step owing to its simple density profile. 
A more complicated particle distribution having a wider dynamic range in the temporal domain of the orbit evolution of individual particles is be a better choice for performance measurements to examine effects of the block time step. 
In order to measure the performance in a realistic distribution, we generate a model of the Andromeda galaxy (\astrobj{M31}). 
The mass distribution model of \astrobj{M31} is given by \citet{Geehan2006, Fardal2007}. 
Its composition is a dark matter halo with an NFW profile (the mass is $8.11 \times 10^{11}$~$M_\odot$ and the scale length is 7.63~kpc) with 7,730,866 particles, a stellar bulge with a Hernquist profile (the mass is $3.24 \times 10^{10}$~$M_\odot$ and the scale radius is 0.61~kpc) with 308,853 particles, and an exponential disk (the mass is $3.66 \times 10^{10}$~$M_\odot$, the scale length is 5.4~kpc, and the scale height is 0.6~kpc) with 348,889 particles. 
The total number of $N$-body particles is $2^{23} =$~8,388,608, the masses of all $N$-body particles are identical and the Plummer softening length is set to 16~pc. 
On M2090, a performance measurement with $\varDelta_\mathrm{mul}$ of $2^{-19}$ for \texttt{GOTHIC} with the multipole MAC and the shared time step was not finished due to the limitation of the execution time on HA-PACS. 
Figure~\ref{fig:elapsed.time@M31} shows the results of the measurements. 
Again, the block time step with the acceleration MAC achieves the best performance in most cases. 
The performance gain of the block time step is significantly greater than that for a pure NFW sphere (Fig.~\ref{fig:elapsed.time@NFW}). 
This is because additional components (the bulge and the disk) make the density profile steeper. 
A steeper density profile means a wider range of time steps of $N$-body particles since the free-fall time, one of the typical time scales of the system, is proportional to the inverse square root of the volume density. 
Indeed, the number of time step levels increases from four for the NFW sphere to five for the \astrobj{M31} model.
The block time step with the acceleration MAC (red filled circles) achieves the shortest elapsed time in most cases, and is always faster than \texttt{Bonsai} (green triangles). 
On M2090, the performance measurement with $\theta$ of 0.1 for \texttt{Bonsai} was not completed due to exceeding the maximum execution time on HA-PACS. 
Since the performance improvements from the shared time step are more significant compared to the pure NFW model, the speed increase of \texttt{GOTHIC} compared to \texttt{Bonsai} is greater in the case of the Andromeda galaxy model compared to the NFW model. 

\subsection{Benefits from Block Time Step}
\label{sec:results:benefits}
\begin{figure*}
  \centering
  \includegraphics[viewport=138 356 830 616, width=.99\linewidth,clip]{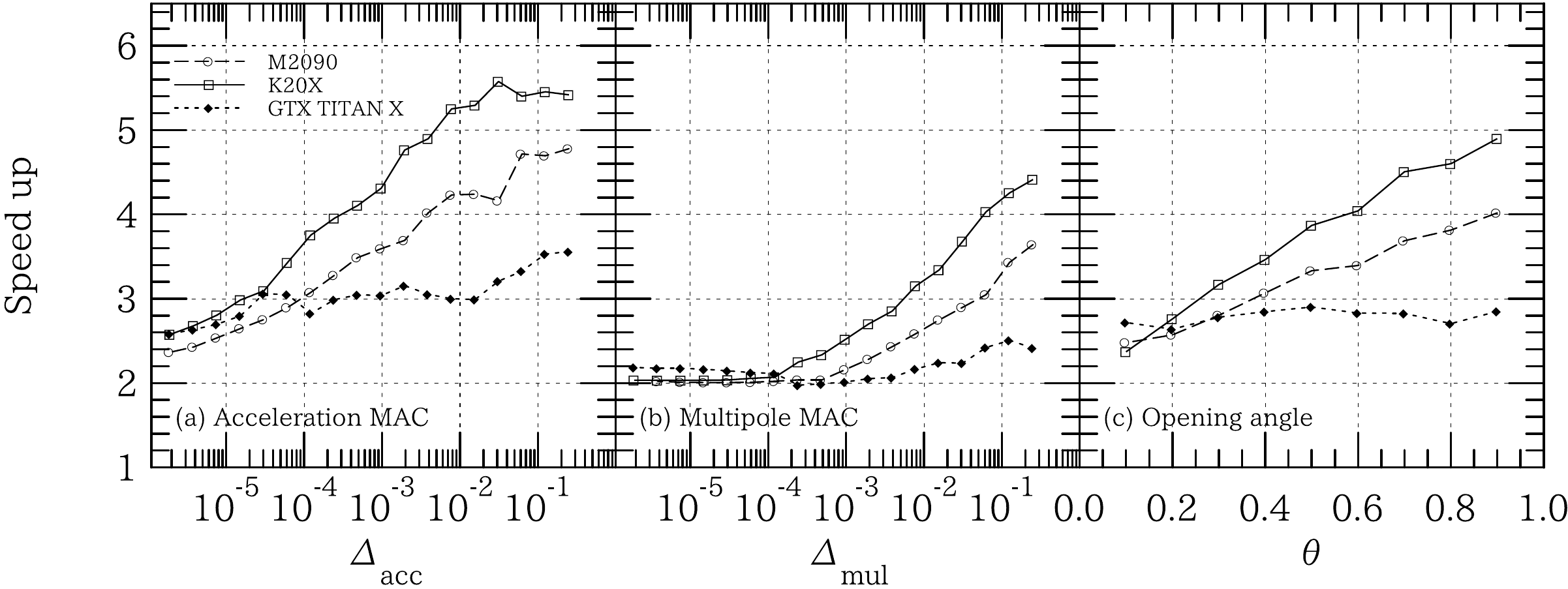}
  \caption{
    Speed up of the block time step compared to the shared time step as a function of the accuracy controlling parameter. 
    The particle distribution is \astrobj{M31} by $2^{23} =$~8,388,608 particles. 
    The open circles with the dashed line, the open squares with the solid line, and the filled diamonds with the dotted line in each panel show the speed up on M2090, K20X, and GTX TITAN X, respectively. 
    Each panels presents different MACs: (a) acceleration MAC \citep{Springel2005}, (b) multipole MAC \citep{WarrenSalmon1993}, and (c) opening angle \citep{BarnesHut1986}. 
  }
  \label{fig:gain@M31}
\end{figure*}
To assess benefits of adopting the block time step in detail, Fig.~\ref{fig:gain@M31} shows the speed up of the block time step from the shared time step in the case of \astrobj{M31}. 
The block time step results in two times faster completion compared to the shared time in all cases. 
In galactic scale $N$-body simulations, the typical value for $\theta$ is 0.5--0.7. 
Corresponding values of $\varDelta_\mathrm{acc}$ and $\varDelta_\mathrm{mul}$ which give similar accuracy are from $2^{-8}$ to $2^{-6}$ and from $2^{-5}$ to $2^{-2}$, respectively (see Fig.~\ref{fig:elapsed.time@M31}). 
For such a typical accuracy, adopting a block time step results in about 2--5 times speed up for all three MACs on M2090, K20X, and GTX TITAN X. 
The amount of speed up tends to improve with increasing values of the accuracy-controlling parameters (i.e., the decreasing of the accuracy). 
When increasing the accuracy of gravity calculations, the number of calculations in high density regions increases because many particles are located near each other. 
Since the speed up of the block time step comes from the reduction of calculations in the low density regions, this increase in calculations weakens the benefits of adopting the block time step. 

\begin{figure*}
  \centering
  \includegraphics[viewport=147 395 926 603, width=.99\linewidth,clip]{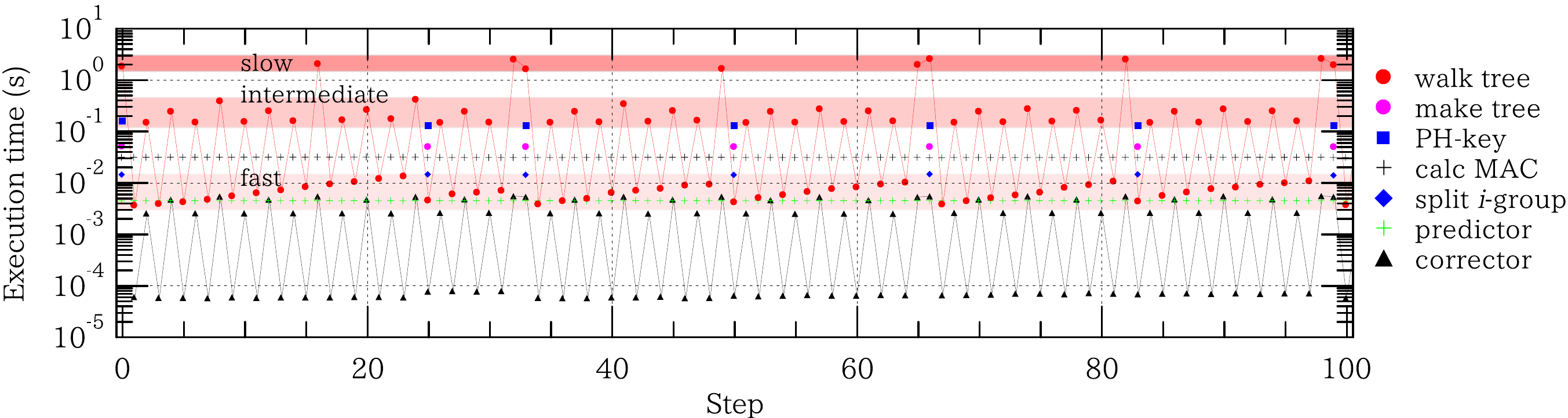}
  \caption{
    Execution time of each function on K20X as a function of the time step. 
    The particle distribution is \astrobj{M31} with $2^{23} =$~8,388,608 particles. 
    The execution time of the function for gravity calculation (red circles connected by red line), tree construction (magenta circles), generation and sorting Peano--Hilbert keys with particles (blue squares), calculating position and mass of pseudo \textit{j}-particles (black crosses), splitting \textit{i}-particles groups (blue diamonds), predicting \textit{j}-particles' position and velocity (green crosses), and correcting velocity of \textit{i}-particles (black triangles connected by black line) are plotted as a function of time steps. 
    The slow, intermediate, and fast sequences are highlighted by bands in three shades of red.
  }
  \label{fig:breakdown@M31}
\end{figure*}
Hereafter, we regard the block time step with the acceleration MAC as a fiducial configuration, and go into more detail about the results from this configuration. 
Figure~\ref{fig:breakdown@M31} shows a breakdown of the execution time of various functions during the first 101 steps of the benchmark with $\varDelta_\mathrm{acc} = 2^{-7} = 7.8125 \times 10^{-3}$ on K20X. 
The initial condition of the system is a model of \astrobj{M31} in dynamical equilibrium with $2^{23} =$~8,388,608 particles. 
A slightly slow execution at the first step pushes back the first tree reconstruction to the 26th step; thereafter, the execution times of all functions settle into a regular repeating pattern because the system is in dynamical equilibrium. 
The execution time for calculating gravity (red circles) is, for the most part, the dominant contribution to the total execution time. 
For the case of the model of \astrobj{M31}, there are three distinct ranges of execution times for calculating gravity; the fast steps with execution times in the range $4 \times 10^{-3}$ s -- $2 \times 10^{-2}$ s, steps with intermediate execution times in the range 0.15 s -- 0.4 s, and the slow steps with execution times of $\sim2$ s). 
We group the steps in these ranges and label them as ``\FSeq'' (\fseq), ``\ISeq'' (\iseq), and ``\SSeq'' (\sseq), respectively. 
The decrease in the number of steps with execution times above 1~s (\FSeq) to ten times during the first 101 steps is a consequence of the block time step reducing the number of calculations for slowly moving \textit{i}-particles. 
This is the main reason for the acceleration by the block time step. 
The achieved mean elapsed time per step is 0.33~s, and is a little above 10\% of the execution time to calculate gravity in the \SSeq. 
The nearly fixed cost to calculate the position and mass of pseudo \textit{j}-particles (black crosses), which is $3.2 \times 10^{-2}$~s, sometimes becomes the most time-consuming function at a given time step. 
This suggests that further optimization of that function might also accelerate the code. 
Performing a more precise time integration is also possible without worsening the total elapsed time. 
For example, one could increase the number of \textit{i}-particles at the cost of an increase in the execution time to calculate gravity.
Unless the increase of the execution time in the \FSeq{} is much greater than that the execution time to update \textit{j}-particles, this would not increase the total elapsed time since the total elapsed time is still dominated by the execution of the \SSeq{} in the gravity calculation is the main reason for the acceleration using the block time step. 
The costs for correcting the velocity of \textit{i}-particles (black triangles) roughly fall into three sequences as well, with execution times of $5 \times 10^{-5}$ s, $2.6 \times 10^{-3}$ s, and $5 \times 10^{-3}$~s.
This implies that the number of \textit{i}-particles within each sequence is fairly constant and suggests the scheme is successfully reducing the calculations of gravity for \textit{i}-particles in the low density regions.
The required time to predict position and velocity of \textit{j}-particles (green crosses) is almost constant at $4.6 \times 10^{-3}$~s, roughly the same as the slowest sequence for the corrector, in every time step. 
This is because the number of \textit{j}-particles is always equal to $N_\mathrm{tot} = 2^{23}$. 

The mean interval between successive tree reconstructions is about 12 steps. 
The costs of functions related to tree reconstruction (generation and sorting Peano--Hilbert keys, sorting $N$-body particles using Peano--Hilbert key, tree construction, and split \textit{i}-particle groups in the low dense region) are almost independent of the particular time step.
Because the radix sorting of 32-bit integers with 64-bit keys, which takes about 0.1~s, is the limiting process, further acceleration of the sorting library is essential to reduce the cost to reconstruct tree structures. 
The execution of the \SSeq{} of the tree traversal and tree reconstruction often form a pair. 
Because tree construction is an order of magnitude faster than tree traversal, even a tiny increase in the cost to traverse the tree structure is greater than the cost of the tree reconstruction, and thus, the execution of the \SSeq{} of tree traversal becomes a trigger to rebuild the tree structure. 

\subsection{Dependence on Number of $N$-body Particles}
\label{sec:results:dependence}
\begin{figure*}
  \centering
  \includegraphics[viewport=138 357 822 630, width=.99\linewidth,clip]{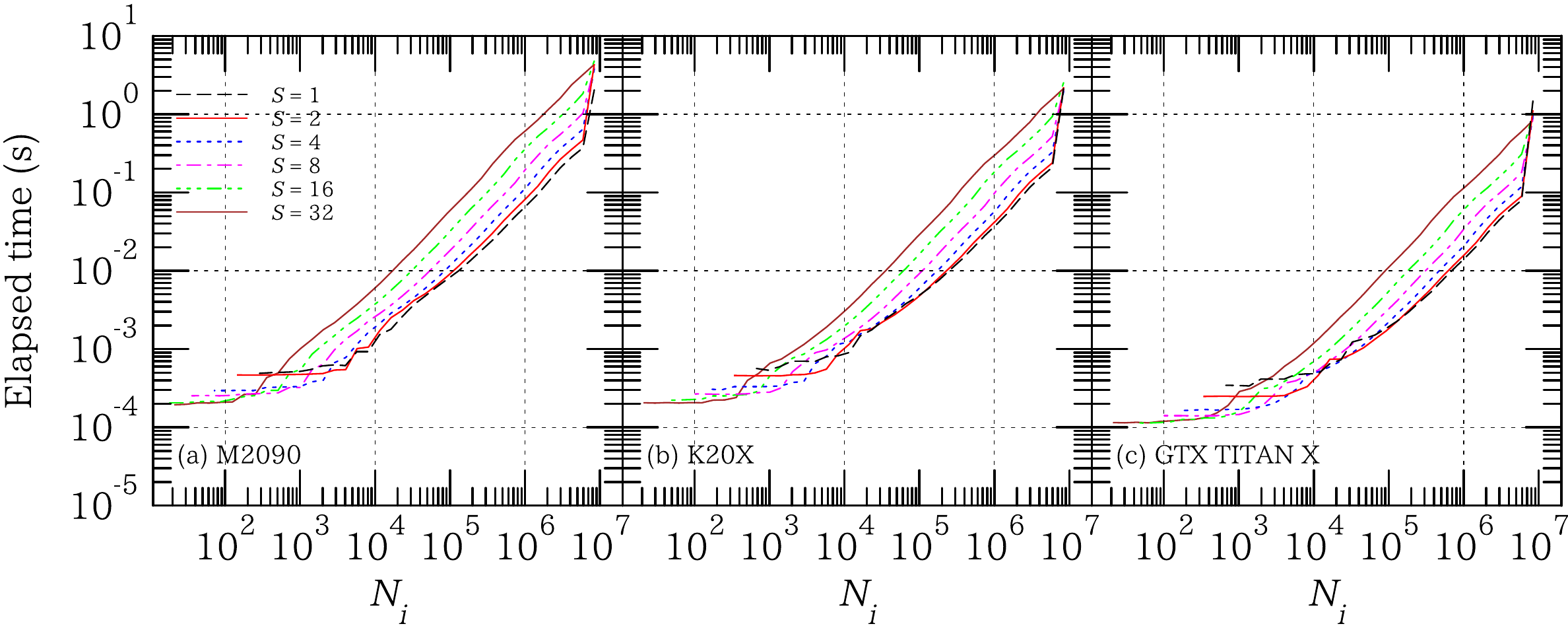}
  \caption{
    Dependence on the number of \textit{i}-particles $N_i$ where the total number of $N$-body particles is $2^{23} =$~8,388,608. 
    The black dashed line ($S = 1$), the red solid line ($S = 2$), the blue dotted line ($S = 4$), the magenta dot-dashed line ($S = 8$), the green triple-dot-dashed line ($S = 16$), and the brown solid ($S = 32$) line represent the elapsed time for the number of threads that share a common \textit{i}-particle $S$. 
    Each panel reveals results on different generation of GPUs: (a) M2090, (b) K20X, and (c) GTX TITAN X. 
  }
  \label{fig:Ni@M31}
\end{figure*}
To examine the effects of \textit{ij}-parallelization, we measured elapsed time while varying the number of \textit{i}-particles, $N_i$, and keeping the total number of $N$-body particles fixed at $N_\mathrm{tot} = 2^{23} =$~8,388,608. 
Figure~\ref{fig:Ni@M31} presents the results for varying number of threads that share a common \textit{i}-particle, $S$, on M2090, K20X, and GTX TITAN X. 
The elapsed time monotonically decreases with $N_i$. 
This feature is associated with the reason for the acceleration by the block time step, and roughly scales as $N_i^1$ if $N_i \gtrsim 10^4 / S$ except for $N_i \sim N_\mathrm{tot}$. 
The steep increase at $N_i \sim N_\mathrm{tot}$ for all cases except for $S = 32$ is related to gravity calculations for \textit{i}-particles in the lowest density regions. 
As noted in \S\ref{sec:implementation:splitGroups}, \texttt{GOTHIC} tends to increase the number of interactions in the low density regions and this causes an increase in the elapsed time. 
Because $T_\mathrm{sub} / S = 32 / S$ particles share the tree traversal, the steepness of the increase becomes weaker with greater $S$ and vanishes for $S = 32$. 
Also, particles in the lowest density regions have the longest free-fall time and would have the longest time step; therefore, they would not be selected as \textit{i}-particles if $N_i < N_\mathrm{tot}$, and this makes the increase of elapsed time steeper.
If further optimizations or another algorithm succeeded in reducing the steep increase of the elapsed time at $N_i \sim N_\mathrm{tot}$, then the total elapsed time \texttt{GOTHIC} could be significantly decreased. 

The critical number $10^4 / S$, which separates the monotonic decrease with $N_i$ and the constant elapsed time irrespective of $N_i$, is determined by the number of running CUDA cores. 
Because the number of thread-blocks per SM is two, the number of threads per block is 256 or 512, and the number of SMs per device is around 20. The number of threads to saturate CUDA cores is given by the product of these three factors and is around $10^4$. 
Introducing \textit{ij}-parallelization activates $S$ times more threads compared to simple \textit{i}-parallelization. 
These two properties result in the critical number being $10^4 / S$. 
The origin and value of the critical number are same for the case of direct summation \citep{Miki2012}. 

\begin{figure*}
  \centering
  \includegraphics[viewport=138 152 821 412, width=.99\linewidth,clip]{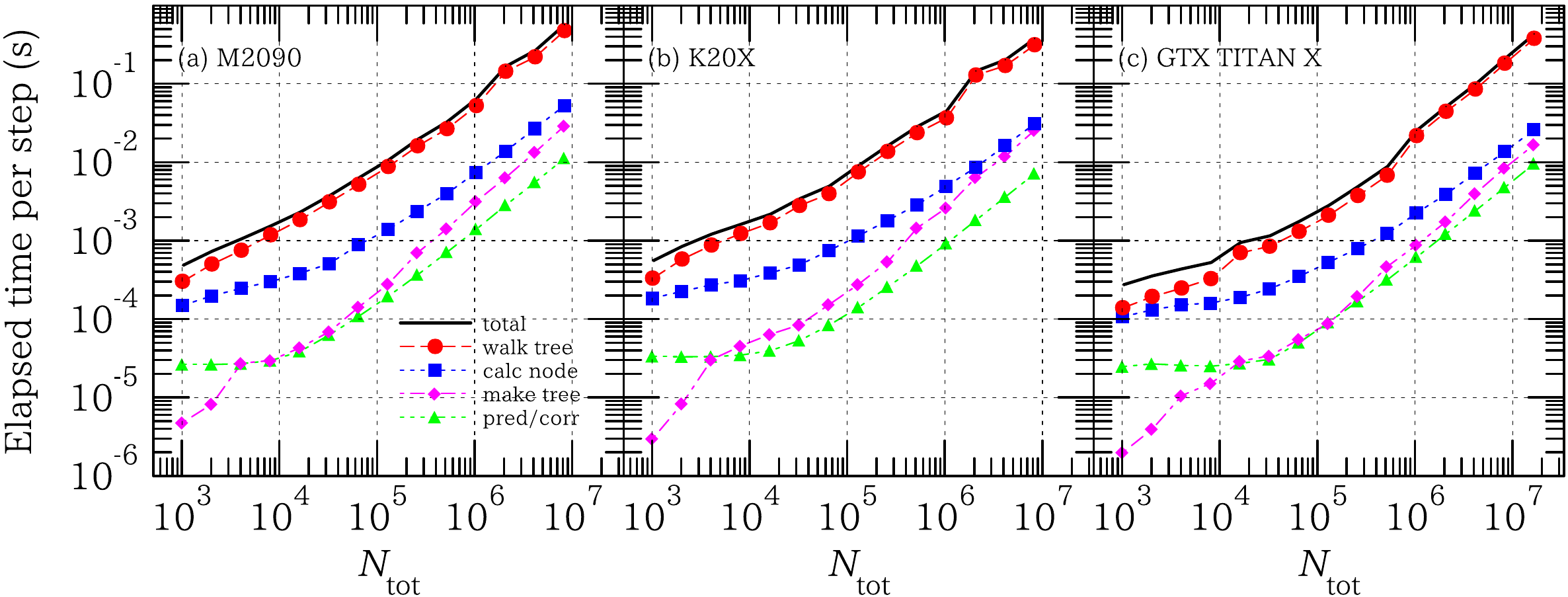}
  \caption{
    Breakdown of the elapsed time of \texttt{GOTHIC} as a function of the total number of $N$-body particles $N_\mathrm{tot}$. 
    Each panel shows the elapsed time of functions for gravity calculation (red circles with dashed line), calculating position and mass of pseudo \textit{j}-particles (blue squares with dotted line), tree construction (magenta diamonds with dot-dashed line), orbit integration (green triangles with triple-dot-dashed line), and sum of them (black solid line). 
    The particle distribution is the \astrobj{M31} model and each panel shows results on different GPUs: (a) M2090, (b) K20X, and (c) GTX TITAN X. 
  }
  \label{Ntot@breakdown}
\end{figure*}
The dependence of \texttt{GOTHIC} on the number of $N$-body particles is the final concern we address. 
Figure~\ref{Ntot@breakdown} presents the elapsed time as a function of the total number of $N$-body particles with $\varDelta_\mathrm{acc} = 2^{-7} = 7.8125 \times 10^{-3}$ on M2090, K20X, and GTX TITAN X. 
Contributions of each function are measured as the elapsed time averaged by 1,024 steps. 
The number of $N$-body particles is changed from $2^{10} =$~1,024 to $2^{24} =$~16,777,216. 
The two-fold greater global memory on GTX TITAN X compared with others enables it to perform $N$-body simulation with $2^{24}$ particles that could not run on M2090 or K20X. 
Traversing the tree structure (red circles with dashed line) always dominates the execution time and scales roughly as $N_\mathrm{tot}$ if $N_\mathrm{tot} \gtrsim 10^5$ on all GPUs. 
It is slightly weaker than the expected scaling of the tree algorithm as $O(N_i \log{N_j})$. 
The scaling gradually becomes worse when decreasing the problem size. 
In $N_\mathrm{tot} \lesssim 10^4$, the execution time to calculate the mass, the position, and the size of pseudo \textit{j}-particles (blue squares with dotted line) approaches a constant floor on each device. 
Furthermore, the floor value is not negligible compared with the elapsed time to calculate gravity and increases its contribution. 
Improving the scaling is also necessary to achieve a shorter time-to-solution for $N_\mathrm{tot} \lesssim 10^4$. 

Contributions from tree construction (magenta diamonds with dot-dashed line) and orbit integration (green triangles with triple-dot-dashed line) are comparable for most values of $N_\mathrm{tot}$ and negligibly small in any case. 
It should be noted that performance optimization of tree construction is also helpful to decrease the time-to-solution even though its execution time itself is negligible. 
As stated in \S\ref{sec:implementation:rebuildInterval}, the interval between successive tree constructions is determined by the balance between execution time of tree traversal and construction. 
Therefore, performance enhancements of the function to update the tree structure can accelerate $N$-body simulation by decreasing the execution time for calculating gravity. 
This is a characteristic behavior of \texttt{GOTHIC} due to optimizations affecting multiple functions. 

The measured elapsed time per step is 0.47~s (0.58~s), 0.39~s (0.38~s), and 0.14~s (0.21~s) for the \astrobj{M31} model (the NFW sphere) with $2^{23} =$~8,388,608 particles on M2090, K20X, and GTX TITAN X, respectively. 
On GTX TITAN X, we ran $N$-body simulation using $2^{24} =$~16,777,216 particles and they took 0.30~s and 0.44~s per step for the \astrobj{M31} model and the NFW sphere, respectively. 
\citet{Ogiya2013} reported that the elapsed time per step of their code was $\sim$5~s on M2090 for the NFW sphere with $2^{24}$ particles. 
This indicates that the sophisticated algorithms and optimizations adopted in \texttt{GOTHIC}, and performance improvements of GPU achieve more than ten times acceleration of $N$-body simulations compared to \citet{Ogiya2013}. 

\section{Discussion}
\label{sec:discussion}
The tree method has a better scaling compared to the direct method and is always faster in the high $N$-regime. 
However, in the low $N$-regime, the direct method becomes faster owing to its simplicity. 
Here, we briefly discuss the transition point at which to switch between the tree method and the direct method. 
\citet{Miki2013} reported that the execution times for calculating gravity by the direct method with $N = 2^{12} =$~4,096 and $N = 2^{13} =$~8,192 on M2090 are $9.7 \times 10^{-4}$~s and $1.9 \times 10^{-3}$~s, respectively. 
They are nearly the same as those with \texttt{GOTHIC} (see Fig.~\ref{Ntot@breakdown}). 
Since $10^4$ is a sufficiently large number of $N$-body particles to obtain the sustained performance on M2090, the growth of the elapsed time is proportional to $N^2$ for $N \gtrsim 10^4$. 
This implies that the tree method becomes faster than the direct method on GPU for $N \gtrsim 10^4$. 
Since direct $N$-body codes on GPU can maintain their $O(N^2)$ scaling down to $\sim 10^3$ through \textit{ij}-parallelization \citep{Miki2012}, direct $N$-body codes becomes faster than the tree method in $N \lesssim 10^4$. 
Furthermore, \citet{Miki2013} adopted the shared time step instead of the block time step, so further speed up of their direct $N$-body code is possible.
In summary, the execution time of \texttt{GOTHIC} is comparable with that of direct $N$-body codes if $N \sim 10^4$ and becomes shorter the larger the problem size. 

\begin{figure*}
  \centering
  \includegraphics[viewport=124 147 820 837, width=.99\linewidth,clip]{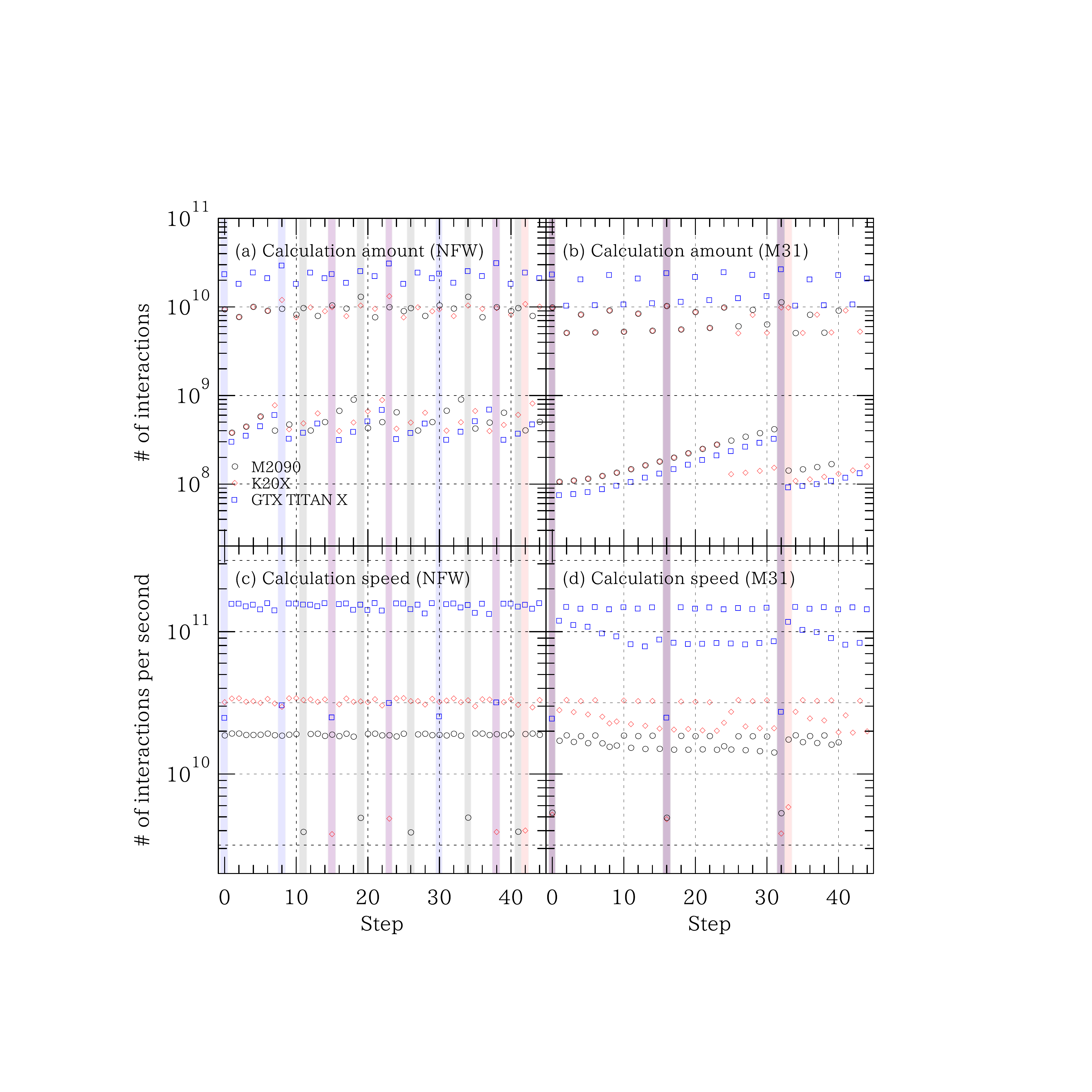}
  \caption{
    Measured performance of \texttt{GOTHIC}. 
    The upper and lower panels present the number of interactions and the calculation speed, respectively, as a function of the time step. 
    Different symbols indicate different GPUs: black circles, red diamonds, and blue squares represent M2090, K20X, and GTX TITAN X, respectively. 
    The left panels show results for the NFW model, and the right ones display results for the \astrobj{M31} model, both with $N = 2^{23} =$~8,388,608. 
    Execution of the \sseq{} is highlighted by vertical bands (colored according to GPU).
  }
  \label{fig:performance}
\end{figure*}
To estimate the achieved performance of \texttt{GOTHIC}, we have first counted the number of interactions computed in each time step. 
The counting of interaction pairs is done in a separate run to that of measurements of the elapsed time in order to remove the additional burden of the performance measurements. 
Figure~\ref{fig:performance} shows the measured results as a function of the time step. 
The directly measured values are the calculated number of interactions in each time step and are shown in Figs.~\ref{fig:performance}a and \ref{fig:performance}b. 
They have similar values on different generations of GPUs. 
The origin of the differences is differences of the configuration of the kernel function to calculate gravitational force (see Tab.~\ref{tab:results:configuration:kernel}). 
The gradual increase in the number of interactions with time step in the \FSeq{} is the reason for the growth of the execution time for calculating gravity while using the same tree structure repeatedly. 
Since rebuilding the tree structure is auto-tuned as described in \S\ref{sec:implementation:rebuildInterval}, the time steps at which the tree is rebuilt will differ depending on the problem or the utilized GPU. 
The number of interactions calculated per second (Figs.~\ref{fig:performance}c and \ref{fig:performance}d) on each GPU is derived by combining independent measurements of the elapsed time. 
The measured results exhibit clear differences among the three GPUs, reflecting their theoretical peak performance. 

The significant difference in each time step is attributable to the block time step. 
Step by step comparison between the number of interactions and the execution time in each time step reveals two things: (1) the lowest calculation speed is associated with the highest number of interaction pairs (as highlighted by vertical bands in Fig.~\ref{fig:performance}) and (2) the minimum number of interaction pairs does not necessarily result in the highest calculation speed (this is more evident in the \astrobj{M31} model). 
The \SSeq{} which corresponds to the maximum number of interaction pairs per step includes all \textit{i}-particles in the lowest density regions, while the \ISeq{} and \FSeq{}, which correspond to the smaller number of interaction pairs per step, do not include \textit{i}-particles in the lowest density regions.
Including \textit{i}-particles in the lowest density regions drastically increases the number of distance evaluations between a group of \textit{i}-particles and pseudo \textit{j}-particles. 
The remedy for this, introduced in \S\ref{sec:implementation:splitGroups}, starts to work at later time steps, and the calculation speed decreases significantly. 
This is also the case with the steep increase of the elapsed time around $N_i \sim N_\mathrm{tot}$ observed in Fig.~\ref{fig:Ni@M31}. 
The lowest number of interaction pairs does not lead to a sustained performance in the \astrobj{M31} model either. 
We find that the highest calculation rate is associated with an intermediate number of interaction pairs. 

\begin{table*}
  \caption{Achieved Performance}
  \label{tab:discussion:performance}
  \centering
  \begin{tabular}{c|c||ccccccc}
    \hline\hline
    GPU & Model$^{(a)}$ & \multicolumn{3}{c}{Number of interactions per second} & \multicolumn{3}{c}{Achieved performance$^{(b)}$ (GFlop/s)} & TPP$^{(c)}$ (GFlop/s)\\
        &              & average & maximum & minimum & average & maximum & minimum & \\
    \hline
    M2090       & NFW & $1.06 \times 10^{10}$ & $1.92 \times 10^{10}$ & $3.87 \times 10^{9 }$ &  296 &  536 & 108 & 1332\\
                & \astrobj{M31} & $1.20 \times 10^{10}$ & $1.86 \times 10^{10}$ & $4.90 \times 10^{9 }$ &  336 &  521 & 137 & \\
    K20X        & NFW & $1.45 \times 10^{10}$ & $3.40 \times 10^{10}$ & $3.77 \times 10^{9 }$ &  377 &  885 &  98 & 3935\\
                & \astrobj{M31} & $1.34 \times 10^{10}$ & $3.30 \times 10^{10}$ & $3.81 \times 10^{9 }$ &  349 &  859 &  99 & \\
    GTX TITAN X & NFW & $6.77 \times 10^{10}$ & $1.59 \times 10^{11}$ & $2.49 \times 10^{10}$ & 1626 & 3827 & 598 & 6611\\
                & \astrobj{M31} & $7.80 \times 10^{10}$ & $1.50 \times 10^{11}$ & $2.46 \times 10^{10}$ & 1871 & 3595 & 590 & \\
    \hline
  \end{tabular}
  \begin{itemize}
    \setlength{\itemsep}{-2pt}
    \item[(a)] Model of initial particle distribution. 
    \item[(b)] One interaction is assumed to correspond to 28, 26 and 24 Flops on M2090, K20X and GTX TITAN X, respectively. 
    \item[(c)] Theoretical peak performance using single precision for each GPU. 
  \end{itemize}
\end{table*}
Conversion from the measured elapsed time to achieved performance requires an assumption about floating-point operation counts per interaction; however, such a conversion is not always rigorous especially in realistic scientific computations. 
Various values of the floating-point operation counts have been adopted in the literature for collisionless $N$-body simulations. 
Examples in studies using GPU(s) are: 20 by \citet{Nyland2007}, 26 by \citet{Miki2012, Miki2013}, and 23 by \citet{Bedorf2014}, while 38 appears to be the typical value used in astrophysics \citep{Kawai1999, HamadaIitaka2007, NitadoriMakino2008, Hamada2009, HamadaNitadori2010, Tanikawa2013}. 
The reason for the differences lies in the estimation of the execution cost of the inverse square root. 
In this study, we assume that the cost of executing the inverse square root corresponds to the ratio of the throughput of the reciprocal square root to that of addition or multiplication. This is found to be 8, 6, and 4~Flops (floating-point operations) on M2090, K20X, and GTX TITAN X, respectively. 
It should be noted that an alternative is adopting 4, 3, and 2~Flops on different generations of GPUs \citep{Capuzzo-DolcettaSpera2013, Bedorf2014}. 
This choice takes into account the fact that GPUs by NVIDIA support FMA operations and thus can execute 2~Flops per clock cycle. 
The remaining operations are three subtractions, three multiplications, and seven FMA operations (20~Flops in total), because \texttt{GOTHIC} calculates not only the gravitational force but also the gravitational potential (an FMA operation returns the potential). 
In summary, we assume that floating-point operation counts per interaction are 28, 26, and 24~Flops, respectively, on M2090, K20X, and GTX TITAN X. 

Table~\ref{tab:discussion:performance} summarizes the measured number of interactions calculated per second and the corresponding performance in units of GFlop/s (Giga Floating-point operations per second) for the NFW sphere and the \astrobj{M31} model with $N = 2^{23} =$~8,388,608 on the three generations of GPUs. 
The averaged performance over time steps on M2090, K20X, and GTX TITAN X are around 320~GFlop/s, 360~GFlop/s, and 1750~GFlop/s, respectively. 
They correspond to 10--30\% of the theoretical peak performance. 
The maximum performance on each GPU is around 40\%, 20\% and 55\% of its theoretical peak performance on M2090, K20X, and GTX TITAN X, respectively. 
Finally, the minimum performance over several time steps drops to less than 10\% of the theoretical peak performance except for the \astrobj{M31} model on M2090. 
This is the case with the highest number of interaction pairs as shown in Fig.~\ref{fig:performance}; i.e., it is equivalent to the performance of the shared time step. 
This means that the benefit of adopting the block time step lies not only in avoiding unnecessary calculations to follow the time evolution of the system but also in increasing the average calculation speed per time step.

\citet{WatanabeNakasato2014} proposed a hybrid tree algorithm to reduce the calculation cost of collisionless $N$-body simulations applying Particle-Particle Particle-Tree (PPPT) algorithm originally developed by \citet{Oshino2011} for collisional systems. 
They divided the gravitational force calculation into two steps, short-range and long-range, and reduce the relative frequency of long-range force calculation. 
Because neglecting small changes of the gravitational field in the distant region does not generate a significant error in the force calculations, they succeeded in accelerating the computations without loss of accuracy. 
They reported a 20\% acceleration of the $N$-body simulation for a Plummer sphere; however, the speed up rate probably depends on the distribution of $N$-body particles. 
In this study, the acceleration by the block time step compared to the shared time step in a Plummer sphere is around 50\% for a given typical accuracy while that in the \astrobj{M31} model reaches 500\%. 
This suggests that the hybrid tree algorithm has the potential to accelerate the calculation more than what was reported by \citet{WatanabeNakasato2014}. 
Also, combining the hybrid tree algorithm with \texttt{GOTHIC} is possible because the original PPPT algorithm was designed to couple with the individual time step scheme. 

There is another unexplored avenue to further accelerate \texttt{GOTHIC}. 
The block time step introduces an order of magnitude variance of the number of \textit{i}-particles $N_i$ in each time step. 
As clearly shown in Fig.~\ref{fig:Ni@M31}, the optimal value for the number of threads to share an \textit{i}-particle, $S$, depends on $N_i$. 
In the current version of \texttt{GOTHIC}, we fix $S$ throughout in the simulation to implement the code easily. 
However, dynamically adjusting the optimal value for $S$ in each time step would accelerate the code especially in the low $N_i$-regime. 
This sort of auto-tuning is suitable to optimize codes whose performance depend strongly on the inputted problems, and might become a key issue to achieve a good strong scaling in future studies. 

Operations for floating point numbers using half precision are supported on current GPUs and are twice as fast as those using single precision on the Pascal generation of GPUs designed for HPC (i.e., GP100 architecture). 
The number of mantissa bits for half precision is 10 in the IEEE 754-2008 standard. 
\citet{Tanikawa2013} showed that the approximate inverse square root function with 12 bits accuracy could provide sufficient accuracy for collisionless systems and implemented this in their software library ``Phantom-GRAPE'', a high-performance direct $N$-body library for CPU. 
This suggests that the approximate inverse square root function using half precision might also give sufficient accuracy for collisionless $N$-body simulations. 
Because the inverse square root is the heaviest function in $N$-body simulations, it would accelerate $N$-body simulations further. 
Even if the accuracy is not sufficient, the Newton--Raphson method can improve the accuracy at only a small cost. 
Furthermore, adopting arithmetic operations using half precision is promising in the tree method since the distance evaluation stage described in \S\ref{sec:implementation:walkTree} does not require a precise value of the distance in single precision. 
Current NVIDIA GPUs support the approximate inverse square root function \texttt{rsqrtf()} with at least 21 bits accuracy \citep{CUDA7.5Manual} for variables at single precision and they were found to successfully accelerate collisionless $N$-body simulations \citep{Nyland2007, Miki2012, Miki2013}. 
Exploiting the half precision version of \texttt{rsqrtf()}, if it exists, would also increase the performance of realistic scientific computations. 

\section{Summary}
\label{sec:summary}
Adopting the tree method is a common way to accelerate collisionless $N$-body simulations in astrophysics, even on GPU. 
Many earlier studies presented tree codes efficiently running on GPU(s), yet none had coupled their code with the block time step \citep{Nakasato2012, Ogiya2013, Bedorf2012, Bedorf2014, WatanabeNakasato2014}. 
Since the block time step can also accelerate $N$-body simulations significantly, we have developed a gravitational octree code (\texttt{GOTHIC}), which is accelerated by the block time step. 
The code adopts the breadth-first search, and runs entirely on GPU, just like \texttt{Bonsai} by \citet{Bedorf2012, Bedorf2014}. 
The algorithm in the tree traversal is an improved version of the algorithm proposed by \citet{Ogiya2013}, which used a depth-first search. 
\texttt{GOTHIC} also does adaptive optimizations, i.e., auto-tuning, by monitoring the execution time of each function. 
The optimizations reduce the time-to-solution by balancing the execution time of multiple functions, and using optional \textit{ij}-parallelization to maintain high performance in the low $N_i$-regime.

The performance of the code is measured on NVIDIA Tesla M2090, K20X, and GeForce GTX TITAN X, which are representative GPUs of the Fermi, Kepler, and Maxwell generation of GPUs, using realistic particle distributions found in astrophysics. 
The results show that the code with the fiducial configuration (the block time step with the acceleration MAC) achieves around a 3--5 times acceleration compared to the shared time step, and is faster than the public code \texttt{Bonsai}. 
The elapsed time of the code scales roughly as $N$ for $N \gtrsim 10^5$; the dependence is slightly weaker than the expected scaling for the tree method, $O(N \log{N})$. 
The averaged performance of the code corresponds to 10--30\% of the theoretical peak performance of each GPU. 
The measured elapsed time per step of \texttt{GOTHIC} is 0.30~s and 0.44~s on GTX TITAN X when the particle distribution represents the Andromeda galaxy and the NFW sphere, respectively, with $2^{24} =$~16,777,216 particles. 
The achieved time-to-solution is more than ten times smaller than that achieved in \citet{Ogiya2013}. 
There are still some possibilities for further optimizations that can be explored, for example: (1) adopting a more sophisticated algorithm such as the hybrid tree algorithm proposed by \citet{WatanabeNakasato2014}, (2) performing deeper optimizations focusing on specific generation of GPUs, (3) auto-tuning of the optimal number of threads $S$ in \textit{ij}-parallelization, and (4) utilizing new functions provided by hardware vendors or compilers such as operations in the half precision. 

\section*{Acknowledgements}
YM is particularly grateful to Go Ogiya for fruitful discussions and providing detailed information about his code implementation and performance measurements. 
YM also appreciates suggestions from Kohji Yoshikawa on code optimizations. 
YM has benefited from feedback by Takanobu Kirihara on using a beta version of \texttt{GOTHIC} which improved the performance in realistic simulations. 
We thank Daisuke Takahashi for providing information on auto-tuning. 
We would like to express our gratitude to Alexander Y.~Wagner for careful reading of the manuscript and comments that improve the paper. 
Numerical simulations were performed on HA-PACS at the Center for Computational Sciences, University of Tsukuba. 
The present study was supported by the Japan Science and Technology Agency's (JST) CREST program entitled ``Research and Development of Unified Environment on Accelerated Computing and Interconnection for Post-Petascale Era.'' 
This research was also supported in part by the Grant-in-Aid for Scientific Research (B) by JSPS (15H03638). 

\appendix
\section{Space-filling Curves}
\label{sec:appendix:space.filling.curves}
\begin{figure*}
\begin{lstlisting}[label=source.code:encPeano, caption=Implementation of Peano--Hilbert key encoder.]
PHint encodePeano3D(int Nlev, PHint px, PHint py, PHint pz){
  PHint key = 0;

  for(int jj = Nlev - 1; jj >= 0; jj--){
    /* get xi, yi, and zi */
    PHint xi = (px >> jj) & 1;
    PHint yi = (py >> jj) & 1;
    PHint zi = (pz >> jj) & 1;

    /* turn px, py, and pz */
    px ^= -( xi & ((!yi) |   zi));
    py ^= -((xi & (  yi  |   zi)) | (yi & (!zi)));
    pz ^= -((xi &  (!yi) & (!zi)) | (yi & (!zi)));

    /* append 3bits to the key */
    key |= ((xi << 2) | ((xi ^ yi) << 1) | ((xi ^ zi) ^ yi)) << (3 * jj);

    /* rotate uncyclic (x->z->y->x) */
    if( zi ){      PHint pt = px;      px = py;      py = pz;      pz = pt;    }
    else{
      /* exchange x and z */
      if( !yi ){   PHint pt = px;      px = pz;                    pz = pt;    }
    }
  }
  return (key);
}
\end{lstlisting}
\end{figure*}
\begin{figure*}
\begin{lstlisting}[label=source.code:decPeano, caption=Implementation of Peano--Hilbert key decoder.]
void decodePeano3D(int Nlev, PHint key, PHint *rx, PHint *ry, PHint *rz){
  PHint px = 0;
  PHint py = 0;
  PHint pz = 0;

  for(int jj = 0; jj < Nlev; jj++){
    /* get xi, yi, and zi */
    PHint xi = (key >> (3 * jj + 2)) & 1;
    PHint yi = (key >> (3 * jj + 1)) & 1;
    PHint zi = (key >> (3 * jj    )) & 1;

    /* rotate cyclic (x->y->z->x) */
    if( yi ^ zi ){
        PHint pt = px; px = pz; pz = py; py = pt; }
    else{
      /* exchange x and z */
      if( (!xi & !yi & !zi) || (xi & yi & zi) ){
        PHint pt = px; px = pz;	         pz = pt; }
    }

    /* turn px , py , and pz */
    PHint mask = ((PHint)1 << jj) - 1;
    px ^= mask & (-( xi & (  yi  |   zi ))                     );
    py ^= mask & (-((xi & ((!yi) | (!zi))) | ((!xi) & yi & zi)));
    pz ^= mask & (-((xi &  (!yi) & (!zi) ) | (        yi & zi)));

    /* append 1 bit to the position */
    px |= ( xi       << jj);
    py |= ((xi ^ yi) << jj);
    pz |= ((yi ^ zi) << jj);
  }
  *rx = px;  *ry = py;  *rz = pz;
}
\end{lstlisting}
\end{figure*}
\begin{figure*}
\begin{lstlisting}[label=source.code:genMorton, caption=Implementation of Morton key generator.]
PHint dilate3D(PHint val){
  val = (val * 0x100000001) & 0x7fff00000000ffff;/*    execute if >= 33 bits */
  val = (val * 0x000010001) & 0x00ff0000ff0000ff;/* 0xff0000ff if <= 30 bits */
  val = (val * 0x000000101) & 0x700f00f00f00f00f;/* 0x0f00f00f if <= 30 bits */
  val = (val * 0x000000011) & 0x30c30c30c30c30c3;/* 0xc30c30c3 if <= 30 bits */
  val = (val * 0x000000005) & 0x1249249249249249;/* 0x49249249 if <= 30 bits */
  return (val);
}
PHint genMorton3D(const PHint ix, const PHint iy, const PHint iz){
  return ((dilate3D(ix) << 2) | (dilate3D(iy) << 1) | (dilate3D(iz)));
}
\end{lstlisting}
\end{figure*}
Listings~\ref{source.code:encPeano} and \ref{source.code:decPeano} are implementations of the Peano--Hilbert key encoder and decoder, respectively, written in C. 
The algorithm is an extension to 3D space of the implementation in 2D space by \citet{LamShapiro1994}. 
The generation of Peano--Hilbert keys boils down to the rotation and/or inversion of the fundamental block. 
Since the rotation and inversion in the 3D space are non-commutative operations, level-by-level encoding/decoding is necessary. 
The number of logical operations is minimized using the Karnaugh map. 
The data type \texttt{PHint} is \texttt{unsigned int} or \texttt{unsigned long int} depending on whether the bit length of the key is less than or equal to 30 (the maximum size that fits in a 32-bit integer), respectively. 

For comparison, Listing~\ref{source.code:genMorton} shows how the Morton key generator works up to 63 bit keys. 
\citet{Bedorf2012} provided Morton key generator in 30 bits based on \citet{RamanWise2008}. 
Listing~\ref{source.code:genMorton} is simply an extension of this to 63 bits. 
It is much simpler than the Peano--Hilbert key generator; however, it does not have a one-stroke sketch nature. 

\section{Comparison of Enclosing Balls}
\label{sec:appendix:enclosing.balls}
We have implemented 5 kinds of enclosing ball generators: (1) the smallest enclosing ball (SEB) given by the algorithm proposed by \citet{Fischer2003}, (2) the efficient bounding sphere (EBS) proposed by \citet{Ritter1990}, (3) the sphere centered on the geometric center of the enclosing rectangular cuboid (GEO), (4) the sphere centered on the center-of-mass of particles (COM), and (5) the smaller of the spheres generated by GEO and COM (CMP). 
The smaller radius of the enclosing ball mitigates the increase of the number of interactions especially in the low density regions and reduce the elapsed time. 
From this point of view, the best choice is the SEB, which has the minimum radius. 
On the other hand, the precise determination of the SEB is a time-consuming process. 
Therefore, the optimal choice for the generator should be determined by comparing the elapsed times of the code with the various generators. 
In this section, we summarize the performance of the enclosing ball generators. 

\begin{figure}
  \centering
  \includegraphics[viewport=124 141 920 835,width=.99\linewidth,clip]{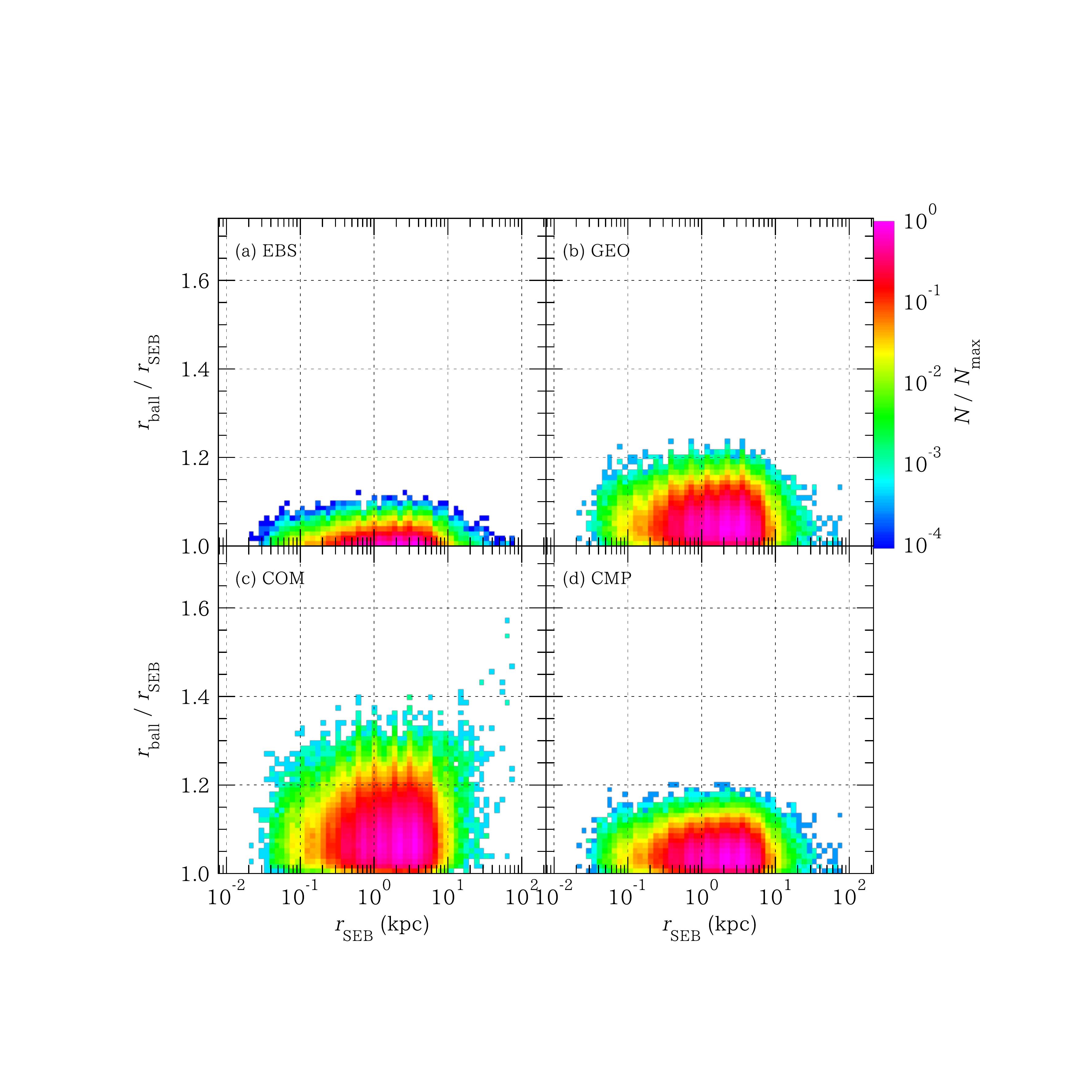}
  \caption{
    Radii of enclosing balls. 
    The horizontal and the vertical axes are the radii of the smallest enclosing ball $r_\mathrm{SEB}$ and that of an enclosing ball $r_\mathrm{ball}$ normalized by $r_\mathrm{SEB}$, respectively. 
    The color map on each panel displays the normalized frequency for different definitions of the pseudo \textit{i}-particles: (a) the efficient bounding sphere \citep{Ritter1990}, (b) the sphere centered on the geometric center of the enclosing rectangular cuboid, (c) the sphere centered on the center-of-mass of particles, and (d) the smaller sphere of (b) and (c). 
   The particle distribution is that representing \astrobj{M31} by $2^{23} =$~8,388,608 particles, and the total number of enclosing balls generated on K20X is 262,144. 
  }
  \label{fig:enclosing.ball}
\end{figure}
First, we compared the radii of each enclosing ball, $r_\mathrm{ball}$. 
Figure~\ref{fig:enclosing.ball} shows amount of radius over-estimation, $r_\mathrm{ball} / r_\mathrm{SEB}$, as a function of the radius of the smallest enclosing ball, $r_\mathrm{SEB}$. 
After SEB, the EBS method results in the smallest radii; its over-estimation is 5\% in most cases and $\sim$ 10\% in the worst case as originally claimed by \citet{Ritter1990}. 
The GEO gives somewhat little bigger radii; however, it is smaller than $1.15 r_\mathrm{SEB}$ in most cases. 
On the other hand, $r_\mathrm{ball}$ in the COM is much bigger, and it exceeds $1.4 r_\mathrm{SEB}$ in the low density regions (i.e., the region with large $r_\mathrm{SEB}$); hence, the number of operations executed in the gravity calculations become much greater than other enclosing ball models. 
The CMP resembles the GEO because the COM predicts larger radii than the GEO in most cases. 

\begin{table*}
  \caption{Computing cost to generate various enclosing balls}
  \label{tab:enclosing.ball}
  \centering
  \begin{tabular}{cc|ccccc}
    \hline\hline
    GPU$^{(a)}$ & Model$^{(b)}$ & SEB$^{(c)}$ & EBS$^{(d)}$ & GEO$^{(e)}$ & COM$^{(f)}$ & CMP$^{(g)}$\\
    \hline
    M2090   & NFW & $2.13 \times 10^{-2}$~s & $1.07 \times 10^{-2}$~s & $5.34 \times 10^{-3}$~s & $3.27 \times 10^{-3}$~s & $7.86 \times 10^{-3}$~s\\
    M2090   & \astrobj{M31} & $2.13 \times 10^{-2}$~s & $1.06 \times 10^{-2}$~s & $5.33 \times 10^{-3}$~s & $3.27 \times 10^{-3}$~s & $7.86 \times 10^{-3}$~s\\
    K20X    & NFW & $1.02 \times 10^{-2}$~s & $3.05 \times 10^{-3}$~s & $1.27 \times 10^{-3}$~s & $9.08 \times 10^{-4}$~s & $2.05 \times 10^{-3}$~s\\
    K20X    & \astrobj{M31} & $1.02 \times 10^{-2}$~s & $3.01 \times 10^{-3}$~s & $1.27 \times 10^{-3}$~s & $9.08 \times 10^{-4}$~s & $2.06 \times 10^{-3}$~s\\
    TITAN X & NFW & $1.06 \times 10^{-2}$~s & $2.04 \times 10^{-3}$~s & $5.09 \times 10^{-6}$~s & $7.45 \times 10^{-4}$~s & $8.43 \times 10^{-4}$~s\\
    TITAN X & \astrobj{M31} & $1.06 \times 10^{-2}$~s & $2.00 \times 10^{-3}$~s & $5.09 \times 10^{-6}$~s & $7.49 \times 10^{-4}$~s & $8.69 \times 10^{-4}$~s\\
    \hline
  \end{tabular}
  \begin{itemize}
    \setlength{\itemsep}{-2pt}
    \item[(a)] Name of GPU. 
    \item[(b)] Particle distribution models. 
    \item[(c)] Cost to generate the smallest enclosing ball based on \citet{Fischer2003}. 
    \item[(d)] Cost to generate the efficient bounding sphere based on \citet{Ritter1990}. 
    \item[(e)] Cost to generate the sphere centered on the geometric center of the enclosing rectangular cuboid. 
    \item[(f)] Cost to generate the sphere centered on the center-of-mass of particles. 
    \item[(g)] Cost to generate the smaller sphere of GEO and COM. 
  \end{itemize}
\end{table*}
Table~\ref{tab:enclosing.ball} lists the costs to generate each enclosing ball on different GPUs. 
The cost is measured by calling the \texttt{clock64()} function within the \texttt{\_\_global\_\_} function in the CUDA code and translated into the elapsed time by dividing by the number of concurrent warps and the clock cycle frequency. 
The elapsed time to generate enclosing balls is always negligibly small compared to that to calculate gravity. 
The dependence of the elapsed time on the particle distribution is much weaker compared to that of the gravity calculation. 

\section{Modeling the Interval of Tree Rebuild}
\label{sec:appendix:rebuild.interval.modeling}
In the power-law growth model, the required time to calculate gravity at the $i$-th step is assumed to grow as 
\begin{equation}
  t_\mathrm{walk}^{(i)} = r^{i - 1} t_1,
\end{equation}
where $t_1$ and $r$ are the scale factor and the common ratio, respectively. 
The total elapsed time after $n$ steps is given by
\begin{equation}
  t_\mathrm{tot} = t_\mathrm{make} + \frac{r^n - 1}{r - 1} t_1.
\end{equation}
The first and the second derivatives of $t_\mathrm{mean} = t_\mathrm{tot} / n$ with respect to $n$ are calculated as 
\begin{align}
  \frac{d}{dn} \frac{t_\mathrm{tot}}{n} &
  = - \frac{t_\mathrm{make}}{n^2} + \frac{(n \ln{r} - 1) r^n + 1}{n^2 (r - 1)} t_1,
  \\
  \frac{d^2}{dn^2} \frac{t_\mathrm{tot}}{n} &
  = \frac{2 t_\mathrm{make}}{n^3} + \frac{\{1 + (n \ln{r} - 1)^2\} r^n - 2}{n^3 (r - 1)} t_1.
  \label{eq:appendix:rebuild.interval.modeling:2nd.derivative.in.power.law.model}
\end{align}
Therefore, the desired condition for rebuilding the tree becomes
\begin{equation}
  (n \ln{r} - 1) r^n = (r - 1) \frac{t_\mathrm{make}}{t_1} - 1,
  \label{eq:appendix:rebuild.interval.modeling:guess.in.power.law.model}
\end{equation}
if the right hand side of (\ref{eq:appendix:rebuild.interval.modeling:2nd.derivative.in.power.law.model}) is positive. 
Substituting (\ref{eq:appendix:rebuild.interval.modeling:guess.in.power.law.model}) into (\ref{eq:appendix:rebuild.interval.modeling:2nd.derivative.in.power.law.model}) yields the equation
\begin{align}
  \frac{d^2}{dn^2} \frac{t_\mathrm{tot}}{n} &
  = \frac{1}{n} \frac{(\ln{r})^2 r^n}{r - 1} t_1,
\end{align}
which implies that $r > 1$ is the necessary condition to minimize $t_\mathrm{mean}$. 

In the parabolic growth model, we assume
\begin{equation}
  t_\mathrm{walk}^{(i)} = t_1 + (i - 1) b + (i - 1)^2 a,
\end{equation}
where $t_1$, $a$, and $b$ are fitting parameters determined by the least squared method. 
The total elapsed time after $n$ steps is written as 
\begin{align}
  t_\mathrm{tot} &
  = t_\mathrm{make} + n t_1 + \frac{n (n - 1)}{2} b + \frac{n (n - 1) (2n - 1)}{6} a.
\end{align}
The first and the second derivatives of $t_\mathrm{mean} = t_\mathrm{tot} / n$ with respect to $n$ are calculated as 
\begin{align}
  \frac{d}{dn} \frac{t_\mathrm{tot}}{n} &
  = - \frac{t_\mathrm{make}}{n^2} + \frac{b}{2} + \frac{4 n - 3}{6} a,
  \label{eq:appendix:rebuild.interval.modeling:1st.derivative.in.parabolic.model}
  \\
  \frac{d^2}{dn^2} \frac{t_\mathrm{tot}}{n} &
  = \frac{2 t_\mathrm{make}}{n^3} + \frac{2 a}{3}.
  \label{eq:appendix:rebuild.interval.modeling:2nd.derivative.in.parabolic.model}
\end{align}
Equating (\ref{eq:appendix:rebuild.interval.modeling:1st.derivative.in.parabolic.model})  to zero yields the optimal choice as
\begin{equation}
  n^2 = \left\{\frac{b}{2} + \frac{4 n - 3}{6} a\right\}^{-1} t_\mathrm{make}. 
  \label{eq:appendix:rebuild.interval.modeling:guess.in.parabolic.model}
\end{equation}
Putting (\ref{eq:appendix:rebuild.interval.modeling:guess.in.parabolic.model}) into (\ref{eq:appendix:rebuild.interval.modeling:2nd.derivative.in.parabolic.model}) gives the expression of the second derivative at the extremum:
\begin{align}
  \frac{d^2}{dn^2} \frac{t_\mathrm{tot}}{n} &
  = \frac{b}{n} + \frac{2 n - 1}{n} a
  = \frac{b + (2 n - 1) a}{n}. 
\end{align}
Therefore, 
\begin{equation}
  (2 n - 1) a + b \geq 0
\end{equation}
is the necessary condition to get the shortest time-to-solution. 

 \bibliographystyle{elsarticle-harv} 
 \bibliography{ref}

\begin{thebibliography}{68}
\expandafter\ifx\csname natexlab\endcsname\relax\def\natexlab#1{#1}\fi
\expandafter\ifx\csname url\endcsname\relax
  \def\url#1{\texttt{#1}}\fi
\expandafter\ifx\csname urlprefix\endcsname\relax\def\urlprefix{URL }\fi

\bibitem[{{Aarseth}(1963)}]{Aarseth1963}
{Aarseth}, S.~J., 1963. {Dynamical evolution of clusters of galaxies, I}.
  \mnras 126, 223.

\bibitem[{Ashari et~al.(2014)Ashari, Sedaghati, Eisenlohr, and
  Sadayappan}]{Ashari2014}
Ashari, A., Sedaghati, N., Eisenlohr, J., Sadayappan, P., 2014. {An efficient
  two-dimensional blocking strategy for sparse matrix-vector multiplication on
  GPUs}. In: Bode, A., Gerndt, M., Stenstr{\"{o}}m, P., Rauchwerger, L.,
  Miller, B.~P., Schulz, M. (Eds.), 2014 International Conference on
  Supercomputing, ICS'14, Muenchen, Germany, June 10-13, 2014. {ACM}, pp.
  273--282.

\bibitem[{{Barnes} and {Hut}(1986)}]{BarnesHut1986}
{Barnes}, J., {Hut}, P., Dec. 1986. {A hierarchical O(N log N)
  force-calculation algorithm}. \nat 324, 446--449.

\bibitem[{{B{\'e}dorf} et~al.(2014){B{\'e}dorf}, {Gaburov}, {Fujii},
  {Nitadori}, {Ishiyama}, and {Portegies Zwart}}]{Bedorf2014}
{B{\'e}dorf}, J., {Gaburov}, E., {Fujii}, M.~S., {Nitadori}, K., {Ishiyama},
  T., {Portegies Zwart}, S., Dec. 2014. {24.77 Pflops on a Gravitational
  Tree-Code to Simulate the Milky Way Galaxy with 18600 GPUs}. ArXiv e-prints.

\bibitem[{{B{\'e}dorf} et~al.(2012){B{\'e}dorf}, {Gaburov}, and {Portegies
  Zwart}}]{Bedorf2012}
{B{\'e}dorf}, J., {Gaburov}, E., {Portegies Zwart}, S., Apr. 2012. {A sparse
  octree gravitational N-body code that runs entirely on the GPU processor}.
  Journal of Computational Physics 231, 2825--2839.

\bibitem[{Bell and Garland(2008)}]{BellGarland2008}
Bell, N., Garland, M., Dec. 2008. {Efficient Sparse Matrix-Vector
  Multiplication on CUDA}. NVIDIA Technical Report NVR-2008-004, NVIDIA
  Corporation.

\bibitem[{Blelloch(1990)}]{Blelloch1990}
Blelloch, G.~E., Nov. 1990. Prefix sums and their applications. Tech. Rep.
  CMU-CS-90-190, School of Computer Science, Carnegie Mellon University.

\bibitem[{{Capuzzo-Dolcetta} and {Spera}(2013)}]{Capuzzo-DolcettaSpera2013}
{Capuzzo-Dolcetta}, R., {Spera}, M., Nov. 2013. {A performance comparison of
  different graphics processing units running direct N-body simulations}.
  Computer Physics Communications 184, 2528--2539.

\bibitem[{{Fardal} et~al.(2007){Fardal}, {Guhathakurta}, {Babul}, and
  {McConnachie}}]{Fardal2007}
{Fardal}, M.~A., {Guhathakurta}, P., {Babul}, A., {McConnachie}, A.~W., Sep.
  2007. {Investigating the Andromeda stream - III. A young shell system in
  M31}. \mnras 380, 15--32.

\bibitem[{Fischer et~al.(2003)Fischer, G{\"{a}}rtner, and Kutz}]{Fischer2003}
Fischer, K., G{\"{a}}rtner, B., Kutz, M., 2003. Fast smallest-enclosing-ball
  computation in high dimensions. In: Battista, G.~D., Zwick, U. (Eds.),
  Algorithms - {ESA} 2003, 11th Annual European Symposium, Budapest, Hungary,
  September 16-19, 2003, Proceedings. Vol. 2832 of Lecture Notes in Computer
  Science. Springer, pp. 630--641.

\bibitem[{Frigo and Johnson(2005)}]{FrigoJohnson2005}
Frigo, M., Johnson, S.~G., 2005. The design and implementation of {FFTW3}.
  Proceedings of the IEEE 93~(2), special issue on "Program Generation,
  Optimization, and Adaptation".

\bibitem[{{Fukushige} et~al.(1991){Fukushige}, {Ito}, {Makino}, {Ebisuzaki},
  {Sugimoto}, and {Umemura}}]{GRAPE1A}
{Fukushige}, T., {Ito}, T., {Makino}, J., {Ebisuzaki}, T., {Sugimoto}, D.,
  {Umemura}, M., Dec. 1991. {GRAPE-1A: Special-Purpose Computer for N-body
  Simulation with a Tree Code}. \pasj 43, 841--858.

\bibitem[{{Fukushige} et~al.(2005){Fukushige}, {Makino}, and {Kawai}}]{GRAPE6A}
{Fukushige}, T., {Makino}, J., {Kawai}, A., Dec. 2005. {GRAPE-6A: A Single-Card
  GRAPE-6 for Parallel PC-GRAPE Cluster Systems}. \pasj 57, 1009--1021.

\bibitem[{{Geehan} et~al.(2006){Geehan}, {Fardal}, {Babul}, and
  {Guhathakurta}}]{Geehan2006}
{Geehan}, J.~J., {Fardal}, M.~A., {Babul}, A., {Guhathakurta}, P., Mar. 2006.
  {Investigating the Andromeda stream - I. Simple analytic bulge-disc-halo
  model for M31}. \mnras 366, 996--1011.

\bibitem[{{Hamada} and {Iitaka}(2007)}]{HamadaIitaka2007}
{Hamada}, T., {Iitaka}, T., Mar. 2007. {The Chamomile Scheme: An Optimized
  Algorithm for N-body simulations on Programmable Graphics Processing Units}.
  ArXiv Astrophysics e-prints.

\bibitem[{Hamada et~al.(2009)Hamada, Narumi, Yokota, Yasuoka, Nitadori, and
  Taiji}]{Hamada2009}
Hamada, T., Narumi, T., Yokota, R., Yasuoka, K., Nitadori, K., Taiji, M., 2009.
  {42 TFlops hierarchical $N$-body simulations on GPUs with applications in
  both astrophysics and turbulence}. In: Proceedings of the Conference on High
  Performance Computing Networking, Storage and Analysis. SC '09. ACM, New
  York, NY, USA, pp. 62:1--62:12.

\bibitem[{Hamada and Nitadori(2010)}]{HamadaNitadori2010}
Hamada, T., Nitadori, K., 2010. {190 TFlops Astrophysical $N$-body Simulation
  on a Cluster of GPUs}. In: Proceedings of the 2010 ACM/IEEE International
  Conference for High Performance Computing, Networking, Storage and Analysis.
  SC '10. IEEE Computer Society, Washington, DC, USA, pp. 1--9.

\bibitem[{{Hernquist}(1990)}]{Hernquist1990}
{Hernquist}, L., Jun. 1990. {An analytical model for spherical galaxies and
  bulges}. \apj 356, 359--364.

\bibitem[{{Hockney} and {Eastwood}(1988)}]{HockneyEastwood1988}
{Hockney}, R.~W., {Eastwood}, J.~W., 1988. {Computer simulation using
  particles}.

\bibitem[{{Ishiyama} et~al.(2009){Ishiyama}, {Fukushige}, and
  {Makino}}]{Ishiyama2009}
{Ishiyama}, T., {Fukushige}, T., {Makino}, J., Dec. 2009. {GreeM: Massively
  Parallel TreePM Code for Large Cosmological N -body Simulations}. \pasj 61,
  1319--1330.

\bibitem[{Ishiyama et~al.(2012)Ishiyama, Nitadori, and Makino}]{Ishiyama2012}
Ishiyama, T., Nitadori, K., Makino, J., 2012. 4.45 pflops astrophysical
  \emph{N}-body simulation on {K} computer: the gravitational trillion-body
  problem. In: Hollingsworth, J.~K. (Ed.), {SC} Conference on High Performance
  Computing Networking, Storage and Analysis, {SC} '12, Salt Lake City, UT,
  {USA} - November 11 - 15, 2012. {IEEE/ACM}, p.~5.

\bibitem[{{Ito} et~al.(1991){Ito}, {Ebisuzaki}, {Makino}, and
  {Sugimoto}}]{GRAPE2}
{Ito}, T., {Ebisuzaki}, T., {Makino}, J., {Sugimoto}, D., Jun. 1991. {A
  Special-Purpose Computer for Gravitational Many-Body Systems: GRAPE-2}. \pasj
  43, 547--555.

\bibitem[{{Ito} et~al.(1990){Ito}, {Makino}, {Ebisuzaki}, and
  {Sugimoto}}]{GRAPE1}
{Ito}, T., {Makino}, J., {Ebisuzaki}, T., {Sugimoto}, D., Sep. 1990. {A
  special-purpose N-body machine GRAPE-1}. Computer Physics Communications 60,
  187--194.

\bibitem[{{Ito} et~al.(1993){Ito}, {Makino}, {Fukushige}, {Ebisuzaki},
  {Okumura}, and {Sugimoto}}]{GRAPE2A}
{Ito}, T., {Makino}, J., {Fukushige}, T., {Ebisuzaki}, T., {Okumura}, S.~K.,
  {Sugimoto}, D., Jun. 1993. {A Special-Purpose Computer for N-Body
  Simulations: GRAPE-2A}. \pasj 45, 339--347.

\bibitem[{Kawai et~al.(1999)Kawai, Fukushige, and Makino}]{Kawai1999}
Kawai, A., Fukushige, T., Makino, J., 1999. {\$7.0/Mflops astrophysical N-body
  simulation with treecode on GRAPE-5}. In: Proceedings of the 1999 ACM/IEEE
  conference on Supercomputing (CDROM). Supercomputing '99. ACM, New York, NY,
  USA.

\bibitem[{{Kawai} et~al.(2000){Kawai}, {Fukushige}, {Makino}, and
  {Taiji}}]{GRAPE5}
{Kawai}, A., {Fukushige}, T., {Makino}, J., {Taiji}, M., Aug. 2000. {GRAPE-5: A
  Special-Purpose Computer for N-Body Simulations}. \pasj 52, 659--676.

\bibitem[{{King}(1966)}]{King1966}
{King}, I.~R., Feb. 1966. {The structure of star clusters. III. Some simple
  dynamical models}. \aj 71, 64.

\bibitem[{Lai and Seznec(2013)}]{LaiSeznec2013}
Lai, J., Seznec, A., 2013. {Performance upper bound analysis and optimization
  of SGEMM on Fermi and Kepler GPUs}. In: Proceedings of the 2013 {IEEE/ACM}
  International Symposium on Code Generation and Optimization, {CGO} 2013,
  Shenzhen, China, February 23-27, 2013. {IEEE} Computer Society, pp.
  4:1--4:10.

\bibitem[{Lam and Shapiro(1994)}]{LamShapiro1994}
Lam, W.~M., Shapiro, J.~M., 1994. {A Class of Fast Algorithms for the
  Peano-Hilbert Space-Filling Curve}. In: Proceedings 1994 International
  Conference on Image Processing, Austin, Texas, USA, November 13-16, 1994.
  {IEEE}, pp. 638--641.

\bibitem[{Liu and Vinter(2015)}]{LiuVinter2015}
Liu, W., Vinter, B., 2015. {CSR5:} an efficient storage format for
  cross-platform sparse matrix-vector multiplication. In: Bhuyan, L.~N., Chong,
  F., Sarkar, V. (Eds.), Proceedings of the 29th {ACM} on International
  Conference on Supercomputing, ICS'15, Newport Beach/Irvine, CA, USA, June 08
  - 11, 2015. {ACM}, pp. 339--350.

\bibitem[{Maggioni and Berger-Wolf(2016)}]{MaggioniBerger-Wolf2016}
Maggioni, M., Berger-Wolf, T., 2016. {Optimization techniques for sparse
  matrix–vector multiplication on GPUs}. Journal of Parallel and Distributed
  Computing 93–94, 66 -- 86.

\bibitem[{{Makino} et~al.(2003){Makino}, {Fukushige}, {Koga}, and
  {Namura}}]{GRAPE6}
{Makino}, J., {Fukushige}, T., {Koga}, M., {Namura}, K., Dec. 2003. {GRAPE-6:
  Massively-Parallel Special-Purpose Computer for Astrophysical Particle
  Simulations}. \pasj 55, 1163--1187.

\bibitem[{{Makino} et~al.(1997){Makino}, {Taiji}, {Ebisuzaki}, and
  {Sugimoto}}]{GRAPE4}
{Makino}, J., {Taiji}, M., {Ebisuzaki}, T., {Sugimoto}, D., May 1997. {GRAPE-4:
  A Massively Parallel Special-Purpose Computer for Collisional N-Body
  Simulations}. \apj 480, 432.

\bibitem[{{McMillan}(1986)}]{McMillan1986}
{McMillan}, S.~L.~W., 1986. {The Vectorization of Small-N Integrators}. In:
  {Hut}, P., {McMillan}, S.~L.~W. (Eds.), The Use of Supercomputers in Stellar
  Dynamics. Vol. 267 of Lecture Notes in Physics, Berlin Springer Verlag. p.
  156.

\bibitem[{{Michie}(1963)}]{Michie1963}
{Michie}, R.~W., 1963. {On the distribution of high energy stars in spherical
  stellar systems}. \mnras 125, 127.

\bibitem[{{Michie} and {Bodenheimer}(1963)}]{MichieBodenheimer1963}
{Michie}, R.~W., {Bodenheimer}, P.~H., 1963. {The dynamics of spherical stellar
  systems, II}. \mnras 126, 269.

\bibitem[{{Miki} et~al.(2012){Miki}, {Takahashi}, and {Mori}}]{Miki2012}
{Miki}, Y., {Takahashi}, D., {Mori}, M., 2012. {A Fast Implementation and
  Performance Analysis of Collisionless N-body Code Based on GPGPU}. Procedia
  Computer Science 9, 96--105, proceedings of the International Conference on
  Computational Science, ICCS 2012.

\bibitem[{{Miki} et~al.(2013){Miki}, {Takahashi}, and {Mori}}]{Miki2013}
{Miki}, Y., {Takahashi}, D., {Mori}, M., Sep. 2013. {Highly scalable
  implementation of an N-body code on a GPU cluster}. Computer Physics
  Communications 184, 2159--2168.

\bibitem[{{Miki} and {Umemura}(in preparation)}]{MAGI}
{Miki}, Y., {Umemura}, M., in preparation. {MAGI: MAny-component Galactic
  Initial-conditions generator}.

\bibitem[{Nakasato(2012)}]{Nakasato2012}
Nakasato, N., 2012. {Implementation of a parallel tree method on a GPU}.
  Journal of Computational Science 3~(3), 132 -- 141, {Scientific Computation
  Methods and Applications}.

\bibitem[{{Navarro} et~al.(1995){Navarro}, {Frenk}, and {White}}]{Navarro1995}
{Navarro}, J.~F., {Frenk}, C.~S., {White}, S.~D.~M., Aug. 1995. {Simulations of
  X-ray clusters}. \mnras 275, 720--740.

\bibitem[{{Navarro} et~al.(1996){Navarro}, {Frenk}, and {White}}]{Navarro1996}
{Navarro}, J.~F., {Frenk}, C.~S., {White}, S.~D.~M., May 1996. {The Structure
  of Cold Dark Matter Halos}. \apj 462, 563.

\bibitem[{{Nelson} et~al.(2009){Nelson}, {Wetzstein}, and {Naab}}]{Nelson2009}
{Nelson}, A.~F., {Wetzstein}, M., {Naab}, T., Oct. 2009. {Vine--A Numerical
  Code for Simulating Astrophysical Systems Using Particles. II. Implementation
  and Performance Characteristics}. \apjs 184, 326--360.

\bibitem[{{Nitadori} and {Makino}(2008)}]{NitadoriMakino2008}
{Nitadori}, K., {Makino}, J., Oct. 2008. {Sixth- and eighth-order Hermite
  integrator for N-body simulations}. \na 13, 498--507.

\bibitem[{{Nitadori} et~al.(2006){Nitadori}, {Makino}, and
  {Abe}}]{Nitadori2006}
{Nitadori}, K., {Makino}, J., {Abe}, G., Jun. 2006. {High-Performance
  Small-Scale Simulation of Star Clusters Evolution on Cray XD1}. ArXiv
  Astrophysics e-prints.

\bibitem[{NVIDIA(2007)}]{CUDA1.0Manual}
NVIDIA, 2007. NVIDIA CUDA Conpute Unified Device Architecture Programming Guide
  Version 1.0.

\bibitem[{NVIDIA(2009)}]{GF110Whitepaper}
NVIDIA, 2009. NVIDIA's Next Generation CUDA Compute Architecture: Fermi.

\bibitem[{NVIDIA(2012)}]{GK110Whitepaper}
NVIDIA, 2012. NVIDIA's Next Generation CUDA Compute Architecture: Kepler GK110.

\bibitem[{NVIDIA(2015)}]{CUDA7.5Manual}
NVIDIA, 2015. CUDA C Programming Guide Version 7.5.

\bibitem[{{Nyland} et~al.(2007){Nyland}, {Harris}, and {Prins}}]{Nyland2007}
{Nyland}, L., {Harris}, M., {Prins}, J., 2007. {Fast $N$-Body Simulation with
  CUDA}.

\bibitem[{{Ogiya} et~al.(2013){Ogiya}, {Mori}, {Miki}, {Boku}, and
  {Nakasato}}]{Ogiya2013}
{Ogiya}, G., {Mori}, M., {Miki}, Y., {Boku}, T., {Nakasato}, N., Aug. 2013.
  {Studying the core-cusp problem in cold dark matter halos using N-body
  simulations on GPU clusters}. Journal of Physics Conference Series 454~(1),
  012014.

\bibitem[{{Okumura} et~al.(1993){Okumura}, {Makino}, {Ebisuzaki}, {Fukushige},
  {Ito}, {Sugimoto}, {Hashimoto}, {Tomida}, and {Miyakawa}}]{GRAPE3}
{Okumura}, S.~K., {Makino}, J., {Ebisuzaki}, T., {Fukushige}, T., {Ito}, T.,
  {Sugimoto}, D., {Hashimoto}, E., {Tomida}, K., {Miyakawa}, N., Jun. 1993.
  {Highly Parallelized Special-Purpose Computer, GRAPE-3}. \pasj 45, 329--338.

\bibitem[{{Oshino} et~al.(2011){Oshino}, {Funato}, and {Makino}}]{Oshino2011}
{Oshino}, S., {Funato}, Y., {Makino}, J., Aug. 2011. {Particle-Particle
  Particle-Tree: A Direct-Tree Hybrid Scheme for Collisional N-Body
  Simulations}. \pasj 63, 881--892.

\bibitem[{{Plummer}(1911)}]{Plummer1911}
{Plummer}, H.~C., Mar. 1911. {On the problem of distribution in globular star
  clusters}. \mnras 71, 460--470.

\bibitem[{Press et~al.(2007)Press, Teukolsky, Vetterling, and
  Flannery}]{Press2007}
Press, W.~H., Teukolsky, S.~A., Vetterling, W.~T., Flannery, B.~P., 2007.
  Numerical Recipes 3rd Edition: The Art of Scientific Computing, 3rd Edition.
  Cambridge University Press.

\bibitem[{Raman and Wise(2008)}]{RamanWise2008}
Raman, R., Wise, D.~S., 2008. Converting to and from dilated integers. {IEEE}
  Trans. Computers 57~(4), 567--573.

\bibitem[{Reguly and Giles(2012)}]{RegulyGiles2012}
Reguly, I., Giles, M., May 2012. Efficient sparse matrix-vector multiplication
  on cache-based gpus. In: Innovative Parallel Computing (InPar), 2012. pp.
  1--12.

\bibitem[{Ritter(1990)}]{Ritter1990}
Ritter, J., 1990. Graphics gems. Academic Press Professional, Inc., San Diego,
  CA, USA, Ch. An Efficient Bounding Sphere, pp. 301--303.

\bibitem[{Sagan(2012)}]{Sagan2012}
Sagan, H., 2012. Space-filling curves. Springer Science \& Business Media.

\bibitem[{{Salmon} and {Warren}(1994)}]{SalmonWarren1994}
{Salmon}, J.~K., {Warren}, M.~S., Mar. 1994. {Skeletons from the treecode
  closet}. Journal of Computational Physics 111, 136--155.

\bibitem[{{Springel}(2005)}]{Springel2005}
{Springel}, V., Dec. 2005. {The cosmological simulation code GADGET-2}. \mnras
  364, 1105--1134.

\bibitem[{{Sugimoto} et~al.(1990){Sugimoto}, {Chikada}, {Makino}, {Ito},
  {Ebisuzaki}, and {Umemura}}]{GRAPE}
{Sugimoto}, D., {Chikada}, Y., {Makino}, J., {Ito}, T., {Ebisuzaki}, T.,
  {Umemura}, M., May 1990. {A special-purpose computer for gravitational
  many-body problems}. \nat 345, 33--35.

\bibitem[{{Tanikawa} et~al.(2013){Tanikawa}, {Yoshikawa}, {Nitadori}, and
  {Okamoto}}]{Tanikawa2013}
{Tanikawa}, A., {Yoshikawa}, K., {Nitadori}, K., {Okamoto}, T., Feb. 2013.
  {Phantom-GRAPE: Numerical software library to accelerate collisionless N-body
  simulation with SIMD instruction set on x86 architecture}. \na 19, 74--88.

\bibitem[{{Umemura} et~al.(2012){Umemura}, {Susa}, {Hasegawa}, {Suwa}, and
  {Semelin}}]{FIRST}
{Umemura}, M., {Susa}, H., {Hasegawa}, K., {Suwa}, T., {Semelin}, B., Oct.
  2012. {Formation and radiative feedback of first objects and first galaxies}.
  Progress of Theoretical and Experimental Physics 2012~(1), 01A306.

\bibitem[{{Warren} and {Salmon}(1993)}]{WarrenSalmon1993}
{Warren}, M.~S., {Salmon}, J.~K., 1993. {A Parallel Hashed Oct-Tree N-Body
  Algorithm}. In: Proceedings of the 1993 ACM/IEEE conference on
  Supercomputing. ACM, pp. 12--21.

\bibitem[{{Watanabe} and {Nakasato}(2014)}]{WatanabeNakasato2014}
{Watanabe}, T., {Nakasato}, N., Jun. 2014. {GPU accelerated Hybrid Tree
  Algorithm for Collision-less N-body Simulations}. ArXiv e-prints.

\bibitem[{Whaley et~al.(2001)Whaley, Petitet, and Dongarra}]{Whaley2001}
Whaley, R.~C., Petitet, A., Dongarra, J., 2001. Automated empirical
  optimizations of software and the {ATLAS} project. Parallel Computing
  27~(1-2), 3--35.

\bibitem[{Xiao and Feng(2010)}]{XiaoFeng2010}
Xiao, S., Feng, W., April 2010. {Inter-block GPU communication via fast barrier
  synchronization}. In: Parallel Distributed Processing (IPDPS), 2010 IEEE
  International Symposium on. pp. 1--12.

\end{thebibliography}





\end{document}